\documentclass[usenatbib, usegraphicx]{mn2e}

\usepackage{array}
\usepackage{amsmath}
\usepackage{times}
\usepackage[page]{appendix}
\newcommand\ion[2]{#1$\;${\scshape{#2}}}             

\begin{document}

\title[Rotation periods of exoplanet host stars]
{Rotation periods of exoplanet host stars}

\author[E. Simpson et al.]
{E. K.~Simpson$^{1,2,3}$\thanks{E-mail: esimpson05@qub.ac.uk},
S. L.~Baliunas$^1$, G. W.~Henry$^4$, C. A.~Watson$^{3}$ \\
$^1$Harvard-Smithsonian Center for Astrophysics, 60 Garden Street, Cambridge, MA 02138, USA\\
$^2$School of Physics \& Astronomy, University of Southampton, Highfield, Southampton SO17 1BJ \\
$^3$Astrophysics Research Centre, School of Mathematics \& Physics, QueenÕs University, University Road, Belfast, BT7 1NN, UK\\
$^4$Center of Excellence in Information Systems, Tennessee State University, Nashville, TN 37209, USA}

\date{Accepted: 21 June 2010}

\maketitle

\begin{abstract}
The stellar rotation periods of ten exoplanet host stars have been determined using newly analysed \ion{Ca}{ii} H \& K flux records from Mount Wilson Observatory and Str\"{o}mgren b, y photometric measurements from Tennessee State University's automatic photometric telescopes (APTs) at Fairborn Observatory. Five of the rotation periods have not previously been reported, with that of  HD 130322 very strongly detected at $P_{rot}=26.1 \pm 3.5$ d. The rotation periods of five other stars have been updated using new data. We use the rotation periods to derive the line-of-sight inclinations of the stellar rotation axes, which may be used to probe theories of planet formation and evolution when combined with the planetary orbital inclination found from other methods. Finally, we estimate the masses of fourteen exoplanets under the assumption that the stellar rotation axis is aligned with the orbital axis. We calculate the mass of HD 92788 b (28 $M_{J}$) to be within the low-mass brown dwarf regime and suggest that this object warrants further investigation to confirm its true nature.
   
\end{abstract}


\section{INTRODUCTION}

The majority of exoplanets have been discovered via the radial velocity technique, in which the orbital inclination cannot be directly measured, so only a lower limit to the planet's mass, $m_{p} \sin{i_{p}}$, can be determined. If we assume that the orbits of exoplanets are generally well aligned with the host star's rotation axis, the true planetary mass can be calculated by substituting the unknown inclination of the planetary orbit ($i_{p}$) with the stellar axial inclination ($i_{*})$.

Is this a fair assumption? In the Solar system, the planets are well aligned with the Sun's rotation axis ($<8$ deg, \citealt{Cox00}) and this is thought to be a result of their condensation from a proto-planetary disc. Therefore, other systems may be expected to show a similar alignment. In the case of transiting planets, the alignment of the stellar rotation and orbital axes in the plane of the sky ($\lambda$) is measurable from spectroscopic observations of the Rossiter McLaughlin (RM) effect \citep{Rossiter24, McLaughlin24} or line-profile tomography techniques \citep{CollierCameron10a} during transit. $\lambda$ has been measured in more than twenty systems and the majority (two-thirds) appear to be well aligned; however, there are several systems which are significantly misaligned, including six which have retrograde orbits: HAT-P-7 b \citep{Winn09c, Narita09}, WASP-8 b \citep{Queloz10}, WASP-33 b \citep{CollierCameron10b} and WASP-2 b, WASP-15 b and WASP-17 b \citep{Triaud10}.

Transiting planets generally represent the population of hot Jupiters that have undergone significant orbital migration and may have experienced violent histories causing the observed misalignments \citep[e.g.,][]{Nagasawa08}. Giant planets with orbital periods on the order of years that have not been significantly affected by orbital migration are more likely to have retained the primordial alignment of the proto-plantary disc and the assumption of alignment may therefore be more applicable. Although it has yet to be shown what proportion of systems are coplanar, it seems reasonable to assume that the majority are well aligned.

The inclination of the stellar rotation axis can be determined by combining the stellar rotation period ($P_{rot}$) with the stellar radius ($R_{*}$) and spectroscopic measurements of the rotational broadening ($v \sin i_{*}$) via: 
\begin{eqnarray}\label{eq1}
\sin i_{*} = v \sin i_{*} \left(\frac { P_{rot} } {2 \, \pi \, R_{*}}\right)
\label{i}
\end{eqnarray}
 
The rotation period of a star may be measured by observing periodicities induced when active magnetic regions move in and out of our line-of-sight. As first observed in the Sun, the emission cores of the \ion{Ca}{ii} H (396.8 nm) and K (393.4 nm) spectral lines brighten in regions of increased magnetic flux. We can measure the temporal variation in stellar magnetic activity from a time series of disk integrated \ion{Ca}{ii} H \& K flux measurements, $S$, (the ratio of fluxes in the H \& K emission cores to those of the nearby continuum, see \citealt{Bal05}). Similarly, disk-integrated visible light reveals periodic variability caused by the passage of surface magnetic features such as starspots \citet{henry95c}. Therefore, stellar rotation, on the order of days for main sequence stars, may be measured from the variations in the \ion{Ca}{ii} H \& K and photometric fluxes.

The presence of magnetic features on a stellar surface can also cause radial velocity and photometric variations that can mask or mimic a planet's orbital or transit signature \citep{SD97,DDB97,Paulson04}. HD 166435 is a prime example; its radial velocity variations were shown to be stellar in nature, explaining the apparent orbital period equalling the stellar rotation period \citep{Queloz01}. The detection of planets around T Tauri stars presents even more of a challenge, where  stellar activity is stronger than in main sequence stars and the absence of a correlation between visible-light radial velocity variations and line bisector span is not sufficient to rule out the presence of starspots as the cause of the variations \citep{Prato08}. 

It is therefore very important to monitor the magnetic activity of exoplanet host stars in order to: a) confirm or refute intrinsic stellar variability as the cause of observations attributed to the presence of exoplanets and b) to find an effective method to remove stellar jitter, allowing us to detect smaller planets and to probe the planet populations around magnetically active stars. Surveys at Mount Wilson and Fairborn Observatories have been making such observations for several decades. 

With additional data, such as from astrometric observations, $i_{p}$ can be found for non-transiting systems  \citep[e.g.,][]{Benedict06,McArthur10} and compared to $i_{*}$ to determine the line-of-sight alignment. In the case of planets that have migrated inwards towards their host stars, the migration mechanism can be probed. Migration theories such as disc-planet interactions \citep{Lin96} are thought not to perturb the orbital inclination and may even drive the system further toward alignment. In contrast, theories involving planet-planet interactions \citep{RF96, WM96} or the Kozai mechanism \citep{Kozai62, WM03} could cause axial misalignment. Measuring the alignment therefore offers a means of discriminating between migration mechanisms. In the case of non-migrating planets, the comparison between $i_{p}$ and $i_{*}$ tells us about the planet formation processes and whether the assumption of coplanarity is correct, because the timescale for coplanarisation is thought to be longer than the main sequence lifetime \citep[e.g.,][]{Winn05,Hale94}.

In this paper we infer the rotation periods of ten exoplanet host stars from periodic variations in \ion{Ca}{ii} H \& K and photometric observations. We then test that the exoplanet inference is not due to magnetic activity. Finally, we calculate the inclination of the stellar rotation axes and, under the assumption of alignment, estimate the planetary mass. 

\section{OBSERVATIONS AND METHOD}

Out of the approximately 350 stars hosting exoplanet companions, \ion{Ca}{ii} H \& K measurements of 57 have been made during the decades of the Mount Wilson HK project, see \citet{Bal05} for details. Denser, intra-seasonal sampling sufficient to reveal rotational modulation, begun in 1980, yields a sample of 36 exoplanet host stars observed on more than 50 occasions. Rotational periodicities in several of these stars have previously been reported \citep{Bal96,Don96,Henry00}. The records were re-calibrated in 2003 to compensate for long term variations in the instruments, standard stars and arc calibration lamp. This, combined with an increased dataset, allowed a search for previously unknown rotation periods.  

In addition, photometric records taken using Tennessee State's T11 0.80-m high-precision automatic photometric telescope (APT) of HD 130322, the most magnetically active (highest log(R'$_{HK}$)) exoplanet host star without an already determined rotation period, were analysed. The  Str\"{o}mgren $b$ and $y$ filter data were combined into a single measurement, $(b+y)/2$, to improve precision to 1 mmag for a nightly observation and 1-2 mmag over a season, see \citet{Henry95b,Henry95a, Henry96, Henry99} for further details of instrumental and data reduction procedures. Nightly observations of the target star and two comparison stars were made between 2002 and 2007 and differential magnitudes calculated.  

A modified Lomb-Scargle periodogram analysis \citep{Scargle82}, with a technique outlined by \citet{HB86} and tailored for the HK database was used to determine periodicities in both datasets. Any seasonal trend was fitted with a low order polynomial and removed from the data and the periodogram analysis applied. The significance of the periodogram peaks was estimated by the false alarm probability, FAP \citep{Scargle82,HB86}. A FAP of 0.1\% is equivalent to a confidence level of 99.9\% that the peak does not arise by chance and is the cut-off used to define the detection of a periodicity. 

Biases can be introduced by time-dependent stellar phenomena such as the growth and decay of active regions. Occasionally, this can affect part of a season so subdivided sections of the seasonal light curve were analysed separately (typically half). The growth and decay of active regions occurs on timescales of $\simeq$ 50--300 d, which can mimic or influence the appearance of a rotational period ($\simeq$ 1--100 d) \citep{Don93,DDB97}. The number of cycles contained in a season is generally small and a transient variation could dramatically alter the power at a particular frequency.  This exemplifies the need to determine the persistence of a period in more than one season. 

\citet{Noyes84} found an empirical relationship between a lower main-sequence star's average magnetic activity, $\langle S \rangle$, and its rotation period, arising from increased magnetic activity induced by rapid rotation. Therefore, from an estimate of surface magnetic activity, the rotation period, $P_{calc}$ of a star can be estimated and compared with the observationally inferred period. The HK project has been monitoring stars for up to four decades and, in many cases, over at least one activity cycle. Therefore the average magnetic activity of the stars in this sample are well known. We have redetermined $P_{calc}$ for these stars using all the available data from the HK records, and we find that they have not changed significantly from the values previously determined from the dataset (e.g., \citealt{Henry00}).

The significance of a determined rotation period is based on three criteria:
\begin{enumerate}
\item The periodogram peak in a season is significant (FAP $< 0.1\%$). 
\item A similar period appears in more than one season.
\item The period conforms with estimates of $P_{calc}$ and is broadly consistent with $v \sin i$ and $R_{*}$ (concidering systematics).
\end{enumerate}

 The determination of the rotation period is qualified by a grade; \textit{confirmed} (periodicities show a very low FAP $<$ 0.0001 and all three criteria are met), \textit{probable} (moderate to low FAP $\sim$ 0.001 and meets most or all of the criteria), \textit{weak} (a weaker FAP $\sim$ 0.01 but meets criteria (i) and (iii)).

\section{RESULTS}

\subsection{Rotation Periods}

\begin{table*}
\centering
\caption{Estimated rotation periods of exoplanet host stars from photometric and \ion{Ca}{ii} H \& K monitoring. 
$N_{seasons}$: Number of seasons for which the star has been monitored and in how many seasons a rotation period was detected.
$P_{rot}$: The range of observed rotation periods.
Grade: The grade designated to the determination of the rotation period. For rotation periods determined in this paper, the grade corresponds to these periods. For stars with only published rotation periods, the grade is the one reported in the literature. 
$ P_{rot}$ $lit$: Previous observationally determined rotation periods from the literature.
$P_{calc}$: Rotation period calculated from $\langle S \rangle$ by the method outlined in \citet{Noyes84}. No value is shown for HD 10697 as it has evolved off the main-sequence and the relationship does not apply. Where $P_{calc}$ or $P_{obs}$ have been reported multiple times using observations from the HK database, the most recent value is shown as it covers the longest timespan of observations and therefore provides the most reliable estimate.} \label{Rot}
\begin{tabular}{lcccccc}
\hline
 &Alternative& &$P_{rot}$   & & $P_{rot}$ $lit$ & $ P_{calc}$  \\
\raisebox{3ex}[0pt]{Star}&Name&\raisebox{3ex}[0pt]{$N_{seasons}$}&(days)&\raisebox{3ex}[0pt]{Grade}&(days)&(days) \\
\hline
\multicolumn{3}{l}{New rotation periods:}&&&&\\
HD 3651	&& 7 of 21 		& $40.2 \pm 4.0$     & \textit{Confirmed}  & $42^{a}$, $44^{b}$                            & $44.1,$ $45.0^{c}$                        \\
HD 9826    &$\upsilon$ And& 1 of 6     	& $7.3 \pm 0.04$               	& \textit{Weak}          & 11--$19^{d}$,  12$^{e}$                                       & $11.8$, $12.0^{c}$                      \\
HD 10697  && 1 of 10   	& $32.6 \pm 1.6$ 	& \textit{Weak}          & $\ldots$                                                 & $\ldots$    	                                   \\
HD 22049   &$\epsilon$ Eri& 9 of 22 	& $11.3 \pm 1.1$ 	& \textit{Confirmed} & 11.1--$12.2^{f}$, 10.0--$12.3^{g}$  & $14.3,$ $17.0^{c}$                 \\
HD 69830    && 1 of 8	 	&  $35.1 \pm 0.8$	& \textit{Probable}    & $\ldots$                                                 & $35.0^{c}$, 35.4                        \\
 HD 89744   && 2 of 22   	& $9.2 \pm 0.7$*	& \textit{Probable}    & $9^{b}$                                                & $11.0^{c},$  $12.9$                                   \\
                      && 4 of 22   	& ($12.8 \pm 0.7$)* & $\ldots$                  & 12.0--$12.6^{h}$                                 & $11.0^{c},$ $12.9$                            \\	    
HD 92788   && 1 of 2		& $22.4 \pm 0.7$ 	& \textit{Weak}          & $\ldots$                                                 & $21.3^{i}$, $31.1$, $32.0^{c}$             \\
HD 130322$\dagger$  && 4 of 6     & $26.1 \pm 3.5$ & \textit{Confirmed} & $\ldots$  						& $8.7^{j}$, 23.5, $30^{c}$               \\
HD 154345  && 2 of 7		&  $27.8 \pm 1.7$     & \textit{Probable}   & $\ldots$				                   & $31.0^{c}$, 31.2    			          \\
HD 217014  && 1 of 22   	& $21.9 \pm 0.4$   & \textit{Probable}   & 21.3--$22.6^{d}$           			 & $29.0^{c}$, $29.5$                            \\
\hline
\multicolumn{3}{l}{Published rotation periods:}&&&\\
HD 38529 &&1 of 2 &...&\textit{Confirmed}&$35.7^{k}$&37.8\\
HD 75732 &55 Cnc&5 of 11&...&\textit{Confirmed}&35--$43^{d}$&44.1\\ 
HD 95128 &47 UMa& 1 of 9&...&\textit{Weak}&$74^{d}$&22.7\\ 
HD 117176  &70 Vir&3 of 9&...&\textit{Weak}&29--$34^{d}$&35.7 \\ 
HD 120136 &$\tau$ Boo&13 of 22&...&\textit{Probable}&2.6--$4.1^{d}$, 3$^{l}$ &7.0\\ 
HD 143761 &$\rho$ CrB&5 of 23&...&\textit{Confirmed}&17--$20^{d}$&21.2\\ 
HD 186427 &16 Cyg B&2 of 11&...&\textit{Weak}&25--$38^{d}$&28.2 \\ 
HD 192263  &&1 of 3&...&\textit{Confirmed}&$24^{m}$&16.9 \\
\hline
\multicolumn{3}{l}{No rotation period detected:}&&&\\
HD 4208  &&0 of 2&...&...&...&24.8 \\
HD 16141  &&0 of 2&...&...&...&15.5\\
HD 37124  &&0 of 4&...&...&...&24.8 \\
HD 50554  &&0 of 1&...&...&...&2.1 \\
HD 52265  &&0 of 3&...&...&...&14.9\\
HD 62509  &&0 of 10&...&...&...&... \\
HD 82943  &&0 of 1&...&...&...&9.8\\
HD 106252 &&0 of 1&...&...&...&24.1 \\
HD 114762  &&0 of 7&...&...&...&7.4 \\
HD 141937  &&0 of 2&...&...&...&19.5 \\
HD 145675  &14 Her&0 of 4&...&...&...&48.8\\
HD 168443  &&0 of 4&...&...&...&37.7\\
HD 177830  &&0 of 2&...&...&...&66.4 \\
HD 187123  &&0 of 4&...&...&...&28.9\\
HD 190360  &&0 of 2&...&...&...&37.4\\
HD 209458  &&0 of 3&...&...&...&18.5 \\
HD 210277  &&0 of 10&...&...&...&40.7\\
HD 217107 &&0 of 4&...&...&...&40.0\\
\hline
\end{tabular}
\begin{flushleft}
\small 
\emph
* HD 89744 shows two discrete periodicities (see  section \ref{HD89733}).\\
$\dagger$ Denotes photometric rather than \ion{Ca}{ii} H \& K observations were used to determine the rotation period.\\
\emph{References:}
$^{a}$\citet{Frick04},
$^{b}$\citet{Bal96},
$^{c}$\citet{Wright04},
$^{d}$\citet{Henry00},
$^{e}$ \citet{Shkolnik08},
$^{f}$\citet{Don96},
$^{g}$\citet{Croll06},
$^{h}$\citet{Noyes84},
$^{i}$\citet{Mayor04},
$^{j}$\citet{Udry00},
$^{k}$\citet{Fischer03}
$^{l}$\citet{Catala07, Donati08}
$^{m}$\citet{Henry02} 
\normalsize
\end{flushleft}
\label{rotation}
\end{table*}

\begin{figure}
\caption{A comparison between the observed range of rotation periods reported here, with values estimated from the empirical relationship of \citet{Noyes84} using HK data and values from the literature. No value is shown for HD 10697 as it has evolved off the main-sequence and the relationship does not apply. }
\centering
\includegraphics[angle=0, width=\linewidth]{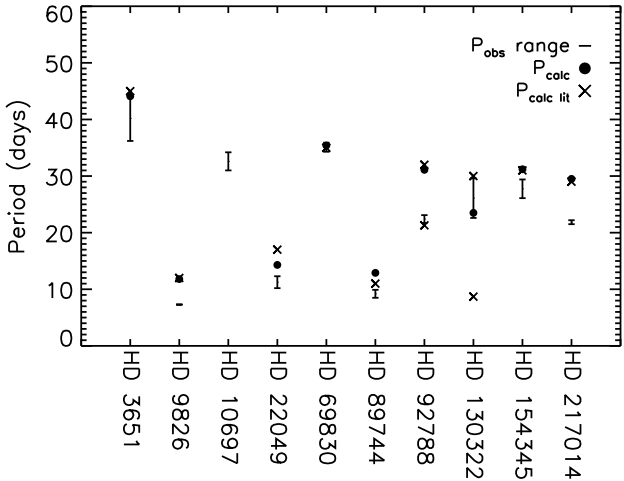} \\[3ex]
\end{figure}

We searched for periodicities associated with stellar rotation in exoplanet host stars in the \ion{Ca}{ii} H \& K records of 36 stars and photometric measurements of HD 130322. Eighteen of these stars showed no variability consistent with rotation (see Table 1). The rotation periods of eight stars in our sample have previously been reported in \citet{Henry00}, \citet{Henry02} and \citet{Fischer03} and re-analysis revealed no new results. In the remaining ten stars, we report new or updated rotation periods. 

Strong and persistent periodicities in multiple observing seasons are seen in three stars: HD 3651, HD 22049 ($\epsilon$\ Eri) and HD 130322 (from photometric data). These stars conform to criteria (i), (ii) and (iii)  so the rotation periods are designated as \textit{confirmed}. Four more stars, HD 69830, HD 89744, HD 154345 and HD 217014 (51 Peg), conform to criteria (i) and (iii) but show moderate periodicities in only one or two seasons and are therefore assigned a grade of \textit{probable}. HD 9826 ($\upsilon$ And), HD 10697 and HD 92788 also conform to criteria (i) and (iii) but show relatively weak periodicities in only one season and the rotation periods are therefore graded as \textit{weak}.

Table \ref{rotation} summarises the rotation periods inferred from the \ion{Ca}{ii} H \& K and photometric records. The range of values likely reflects the uncertainty in the determination of the rotation periods caused by limited data coverage over few rotation cycles in each season, coupled with continual active-region growth and decay rather than a detection of differential rotation. However, differential rotation has been observed in several exoplanet host stars, including $\tau$ Boo \citep{Catala07, Donati08}, which has a rotation period of 3.0 d and 3.9 d at the equator and pole respectively. Differential rotation can been measured using several techniques, including doppler imaging of fast rotating stars (e.g., AB Dor, \citet{Cameron02} and spot modelling of high precision photometric observations (e.g., $\kappa^{1} Ceti$ \citet{Walker07}). For  stars with rotation periods of 5-10 d, the difference between the rotation period of the equator and pole is around 0.5 d, and for slower rotating stars ($\sim$40 d), is around 10 d  \citep{Barnes05}. Thus for the stars in our sample for which we have multiple seasons of data, the error bars on the determination of the rotation period are on the order that we expect to arise from spots at different latitudes. 

The lightcurves and periodograms for each star and season are shown in the appendix. The vertical dashed lines in the periodograms indicate the orbital period of the planet(s). The longer orbital periods do not appear on the plot as the orbital frequency is close to zero on the scale. The dotted horizontal lines show the cut-off for a significant detection of a period, FAP = 0.1. Occasionally there is a second significant peak in the periodogram. In the majority of cases, this disappeared when the primary peak was removed and is therefore likely to be a harmonic of the main peak. Those cases where this does not occur are mentioned in the text. We removed the rotation modulation from each season separately, and searched for other significant periodicities, especially at the orbital period of the planet, however none were found.

\subsection{Analysis of Individual Stars}\label{stars}

\subsubsection{HD 3651}
The star HD 3651 appears to be entering a Maunder-minimum phase as it displays an inter-decadal decrease in magnetic activity with a superimposed decadal cycle whose amplitude has been decreasing with time \citep{Don95}. The most significant periodicities are seen at its most recent maximum phase of activity cycle (1992--1996, $S \simeq 0.18$ ). The previous cycle maximum (1976--1980) had a higher magnetic activity ($S \simeq 0.2$) and may have also shown significant periodicity on a rotational timescale; however, the \ion{Ca}{ii} H \& K records of that period are too sparse to reveal this. The increased HK dataset has expanded the determination of the rotation period from 42 d \citep{Bal96} to $40.2 \pm 4.0$ d. The presence of significant, multiple periodicities, in agreement with values of $P_{calc}$, earns the rotation period a grade of \textit{confirmed}.  

\subsubsection{HD 9826 ($\upsilon$ And)}
The HK project began observations of HD 9826 after the announcement of the discovery of a candidate exoplanet companion in 1996 \citet{Butler97}, therefore there are only six years of data available. HD 9826 is estimated from $\langle S \rangle$ to have a rotation period of $\simeq$ 12 d (\citealt{Wright04} and HK records). \citet{Henry00} report low amplitude periodicities in the \ion{Ca}{ii} H \& K records of 11 d in Season 0 and 19 d in Season 2. We do not find the 11 d period in the re-analysed data of Season 0 but the periodicity in Season 2 does re-occur although at a slightly shorter value of 18 d. However, based on $v \sin i = 9.6 \pm 0.5$ km s$^{-1}$ \citep{VF05} and $R_* = 1.64_{-0.05}^{+0.04}  \,\text{R}_\odot$ \citep{T07}, rotation periods greater than 8 d yield the unphysical result of $\sin i > 1$. We conclude that the 18 d period is unlikely to be the result of rotation as it is inconsistent with $R_{*}$ and $v\sin i$. It may however be a beat period of the two periods in the system, P$_{orb}$ and P$_{rot}$. Upon re-examination of the data, a weak period is seen at 7 d in Season 3. This periodicity is not strong nor repeated in other seasons so cannot be confirmed as rotation and has been assigned a grade of \textit{weak}. \citet{Shkolnik05, Shkolnik08} observed a possible on/off synchronisation between the stellar rotation and the planetary orbital period. At times, the Ca II observations appear to phase with the orbital period suggesting planet-induced activity, and at others, favour a period of $\sim$ 12 d, suggesting this to be the rotation period.
  
\subsubsection{HD 10697}
The rotation period of HD 10697 has not previously been observed and cannot be estimated using the method outlined by \citep{Noyes84} because the relationship is valid only for main-sequence stars and HD 10697 has evolved into a sub-giant. However if we apply the relationship anyway, we obtain $P_{calc}$ $\sim$ 35 d. Initial analysis showed no strong periodicities; however, a review of the light curves by eye showed a steep trend in Season 6 which, when removed, reveals a weak period of 33 d. The other seasons are too sparsely sampled to reveal reliable periods. The period is therefore assigned a grade of \textit{weak}.  

\subsubsection{HD 22049 ({$\epsilon$ Eri})}
HD 22049 has a well-defined rotation period, owing to over 5000 observations in 22 seasons. It is a young, active star ($<1$ Gyr, $\langle S \rangle \simeq 0.5$) and as a result, strong periodicities consistent with rotation are visible in around half of the seasons. Rotational modulation can be seen in almost every season with enough data but factors such as the presence of multiple active regions can cause no clear period to be determined within a season. During Season 0, the periodogram shows two strong, close periods showing that there are two likely solutions due to the short temporal coverage, so we use the strongest peak.

The strong and persistent periodicities seen in the \ion{Ca}{ii} H \& K records confirm the rotation period of HD 22049 a 11.3$ \pm $1.1 d. This updates the value of 11.1--12.2 d found in the \ion{Ca}{ii} H \& H records by \citet{Don96}. More recent, high precision photometry from the \textit{MOST} satellite found a rotation period of 10.0--12.3 d \citep{Croll06} which agrees very well with the values found from the \ion{Ca}{ii} H \& K data and $P_{calc}$ from the HK data. \citet{Wright04} estimate $P_{calc} = 17$ d; however, their observations of $S$ may have not been made over a long enough timespan to obtain a robust $\langle S \rangle$ in the highly variable HD 22049, which may account for the discrepancy. A grade of \textit{confirmed} is assigned.

\subsubsection{HD 69830}
The rotation period of HD 69830 has not previously been observed but $P_{calc}$ is estimated to be $\simeq$ 35 d (\citealt{Wright04} and HK records). The star has been monitored since 1992; however, only Season 1 shows significant periodicity consistent with $P_{calc}$. The 35.1 d periodicity is strong but does not occur in other seasons so a grade of \textit{probable} is assigned. 

\subsubsection{HD 89744}\label{HD89733}
HD 89744 appears to have two discrete periodicities near the value of $P_{calc}$. Season 7 shows two strong peaks in the periodogram, 9.8 d and 12.8 d. Both periodicities have previously been seen in the HK data and reported in the literature as rotation \citep{Bal96,Noyes84} and match well with the values of $P_{calc}$. However, the inclination determined using the longer period ($12.8 \pm 0.7$) gives an unphysical $\sin i > 1$ for $v \sin i = 9.5 \pm 0.5$ km s$^{-1}$ \citep{VF05} and $R_* = 2.14_{-0.20}^{+0.11}\text{R}_\odot$ \citep{T07}. This suggests that the true rotation period is the shorter of the two observed ($9.2 \pm 0.7$ d, assuming $R_{*}$ and $v \sin i$ are reliable) . The shorter periodicity is listed as \textit{probable} and the longer periodicity, while the peaks are statistically significant, is listed in parentheses in Table \ref{Rot} and is likely due to the uncertainty in the determination of the rotation period due to limited temporal coverage and growth and decay of active regions. 

\subsubsection{HD 92788}
There is a discrepancy in values of $P_{calc}$: \citet{Mayor04} calculate 21.3 d whereas \citet{Wright04} and the HK records show a value of 32.0 d. A weak period of 22.4 d is seen in Season 1, which matches well with the period estimated by \citet{Mayor04}; however, the period does not reoccur so earns a grade of \textit{weak}. 

\subsubsection{HD 130322}
The rotation period of HD 130322 has not previously been reported, but is a promising candidate for modulation observations because of its high magnetic activity and variability. All seasons of \ion{Ca}{ii} H \& K flux have less than 35 observations and no significant periodicities can be seen. Differential photometry with dense seasonal coverage reveals strong periodicities in two-thirds of the seasons. The data, which are taken with respect to each of two different comparison stars, show almost identical periods and suggests that they arise from HD 130322 rather than either comparison star.

There is a wide discrepancy in the reported values of $P_{calc}$ estimated for HD 130322: 8.7 d is calculated by \citet{Udry00} whereas \citet{Wright04} report 30.0 d, much closer to the observed periodicity of $26.1 \pm 3.5$ d seen here. The value calculated from the HK observations, 23.5 d, also agrees well with periodicities seen in the photometric records. The coverage of the data used by \citet{Udry00} may have been too short to determine a reliable value of $\langle S \rangle$. The strong and reoccurring periodicities plus agreement with $v \sin i$ and $P_{calc}$ earn a grade of \textit{confirmed}. 

\subsubsection{HD 154345}

There are eight seasons of observations of HD 154345; however, they are all sparsely sampled with less than 50 data points. Despite this, two seasons show periodicities close to $P_{calc} \simeq$ 31 d (\citealt{Wright04} and HK records). There is a weak, 26.8 d periodicity in Season 2 and a prominent period of 29.0 d  in Season 3. Both periodicities are listed and the determination is designated the grade of \textit{probable}.

\subsubsection{HD 217014 ({51 Peg})}
Weak periodicities, $\simeq$ 22--23 d, have previously been reported in the HK observations by \citet{Henry00}. Re-analysis reveals the periods in 1980 and 1984 are likely to be spurious but the 22 d period in 1998 is stronger than previously reported. The reported values of $P_{calc}$ for HD 217014 are close; 29.0 and 29.7 d \citep{Wright04,Henry00}. The discrepancy between $P_{calc}$ and $P_{rot}$ may be due to a lack of long term monitoring to obtain a reliable estimate of  $\langle S \rangle$. The grade is elevated from $weak$ to $probable$. 

\subsection{Orbit and stellar activity}\label{activity}
  
One important support for the inference of a planet from radial velocity variations occurs when $P_{rot}$ and $P_{orbit}$ are dissimilar. The inferred rotation period of HD 69830 (35 d) is similar to the orbital period of the second planet in the system (32 d), however \citet{Lovis06} found no correlation between bisector variations and stellar rotation period so this scenario is unlikely. None of the other stellar rotation periods reported here match the orbital periods of their companions, nor were any significant periodicities close to the planetary orbital periods detected in any season for either close-in or far out planets. This suggests that the radial velocity variations attributed to an orbiting planet are not due to stellar activity and supports the inference of planetary companions in these cases.

Some short period planets show star-planet interactions, including HD 9826, which exhibits intermittent planet-induced magnetic activity close to the orbital period of the planet \citep{Shkolnik05}. We only observed a periodicity in one season (7.3 d in 1999), and although this is close to the orbital period of the planet (4.6 d), it is unlikely to be attributed to planet-induced activity and is more likely the rotation period of the star. An observing season typically spans approximately 150 days with three data points taken each night, on average every 5 nights. Therefore it was not possible to detect intermittent star-planet interactions due to the low time sampling and S/N ($~$50) of the observations.

\subsection{Planetary inclination \& mass} \label{IM}

\begin{table*}
\centering
\setlength{\extrarowheight}{0.2cm}
\caption{Properties of exoplanets and their host stars. Column 3: Exoplanet orbital period. Column 4: Rotational broadening from \citet{VF05}. Column 5: Stellar radius from \citet{T07}. Column 6: Inclination of the stellar rotation axis calculated using $v \sin i$, $R_{*}$ and the stellar rotation period presented in Table \ref{Rot}, using Equation 1. Column 7: Minimum planet mass obtained from radial velocity measurements \citep{Butler06}. Column 8: Mass range calculated by removing the $\sin i$ ambiguity from the minimum planet mass using the inclination of the stellar rotation axis, assuming this is aligned with the planetary orbital axis. For cases where there is no upper mass limit, no constraint can be placed as the minimum inclination is zero.} \label{incmass}	
\begin{tabular}{lccccccc}
\hline
Planet&Alternative Name &P$_{orbit}$ (d)& $v \sin i$ (km s$^{-1})$&$R_*$ ($\text{R}_\odot$)&$i_{*}$ (deg)&$m_p \sin i$ ($M_{J})$ & $m_{p} (M_{J}$) \\[1ex]
\hline
HD 3651 b 	&...&62.23$^{a}$& $1.1 \pm 0.5$ 	& $0.88_{-0.02}^{+0.03}$  	  & $83_{-56}^{+7}$	 &$0.23\pm 0.023$		& $0.2_{-0.0}^{+0.3}$\\
HD 9826 b	&$\upsilon$ And b&4.617$^{b}$& $9.6 \pm 0.5$	& $1.64_{-0.05}^{+0.04}$    	 & $58_{-7}^{+9}$	 &$0.69\pm 0.058$		&$0.8_{-0.1}^{+0.2}$ \\ 
HD 9826 c	&$\upsilon$ And c&240.9$^{b}$&$9.6 \pm 0.5$	&  $1.64_{-0.05}^{+0.04}$   	 & $58_{-7}^{+9}$	 &$1.98\pm 0.170$		&$2.3_{-0.3}^{+0.5}$ \\
HD 9826 d	&$\upsilon$ And d&1282$^{b}$&$9.6 \pm 0.5$	& $1.64_{-0.05}^{+0.04}$   	 & $58_{-7}^{+9}$	 &$3.95\pm 0.330$		&$4.7_{-0.8}^{+0.8}$ \\
HD 10697 b 	&... &1072.3$^{c}$ & $2.5 \pm 0.5$	&$1.73_{-0.07}^{+0.06}$	  & $69_{-26}^{+21}$ &$6.38\pm 0.530$		& $6.9_{-1}^{+3}$ \\
HD 22049 b 	&$\epsilon$ Eri b& 2502$^{d}$ &$2.4 \pm 0.5$	&$0.77_{-0.01}^{+0.02}$	  & $44_{-15}^{+24}$ &$1.06\pm 0.160$		& $1.5_{-0.5}^{+2.5}$ \\
HD 69830 b 	&...&8.667$^{e}$& $0.3 \pm 0.5$	&$0.90_{-0.02}^{+0.02}$	  & $13_{-13}^{+27}$ &$0.03\pm 0.001$		& $0.13_{-0.09}$\\
HD 69830 c	&...&31.56$^{e}$&$0.3 \pm 0.5$		&$0.90_{-0.02}^{+0.02}$	  & $13_{-13}^{+27}$ &$0.04\pm 0.002$		&$0.17_{-0.11}$ \\
HD 69830 d	&...&197$^{e}$&$0.3 \pm 0.5$		&$0.90_{-0.02}^{+0.02}$	  & $13_{-13}^{+27}$ &$0.06\pm 0.004$		& $0.26_{-0.17}$\\
HD 89744 b 	&...&256.0$^{f}$& $9.5 \pm 0.5$	&$2.14_{-0.20}^{+0.11}$	  & $54_{-12}^{+36}$ &$8.58\pm 0.710$		& $11_{-3}^{+4}  $\\
HD 92788 b 	&...& 326.7$^{g}$&$0.3 \pm 0.5$	&$1.00_{-0.03}^{+0.03}$	  & $8_{-8}^{+14}$ 	 &$3.67\pm 0.300$		& $28_{-19}$ \\
HD 130322 b	&...& 10.72$^{h}$&$1.6 \pm 0.5$	&$0.85_{-0.03}^{+0.03}$	  & $76_{-42}^{+14}$ &$1.09\pm 0.098$		& $1.1_{-0.1}^{+1.0}$\\
HD 154345 b	&...& 3340$^{i}$&$1.2 \pm 0.5$	&$0.86_{-0.02}^{+0.03}$	  & $50_{-26}^{+40}$ &$0.95\pm 0.090^{i}$		& $1.2_{-0.4}^{+1.3}$ \\
HD 217014 b	&51 Peg b&4.23$^{j}$& $2.6 \pm 0.5$	&$1.15_{-0.04}^{+0.04}$	 & $79_{-30}^{+11}$	 &$0.47\pm0.039$		& $0.5_{-0.1}^{+0.2}$\\
\hline
								    
\end{tabular}
\begin{flushleft}
$^{a}$ \citet{Fischer03b}, $^{b}$ \citet{McArthur10}, $^{c}$ \citet{Vogt00}, $^{d}$ \citet{Benedict06}, $^{e}$ \citet{Lovis06}, $^{f}$ \citet{Korzennik00}, $^{g}$ \citet{Fischer01}, $^{h}$ \citet{Udry00}, $^{i}$ \citet{Wright08}, $^{j}$ \citet{MQ95}
\end{flushleft}							    									
\end{table*}

The stellar axial inclination was calculated through Equation \ref{eq1} using the values of $v \sin i$ and $R_*$ presented in Table \ref{incmass}. Values of $i$ where $\sin i > 1$, are rejected as unphysical. We assume that the observed rotation period originated from low latitudes and corresponds to the equatorial rotation period. If, due to differential rotation, the spot modulation we observe arises from higher latitudes we would overestimate the equatorial rotation period and from equation \ref{eq1}, slightly overestimating the stellar inclination and thus underestimating the planet mass.

\citet{Benedict06} used astrometric observations to tightly constrain the orbital parameters of HD 22049 b ($i_{p}=30\pm3.8$ deg). The inclination of the planetary orbit lies within the range estimated here for $i_{*}$ ($44_{-15}^{+24}$ deg) and the best-fitting stellar axial inclination (27--33 deg) determined by \citet{Croll06} from the modelling of spots. Also, a circumstellar disc with $i$=25 deg was found by \citet{Greaves98,Greaves05}. All these values suggest that the system is well aligned. The stars with previously published rotation periods (second section of Table \ref{Rot}) are analysed in \citet{Watson10}.

We removed the $\sin{i}$ ambiguity from the planetary minimum mass, $m_{p}\sin{i}$, using the inclination of the stellar rotation axis, under the assumption that the rotation axis is aligned with the planetary orbital axis. Several systems contain multiple planets allowing the mass of fourteen planets to be calculated.  The majority of the planets are calculated to have inclinations greater than 30 deg so the planetary masses are not more than twice the minimum mass. All but one of the planets (HD 92788) have calculated masses below the brown-dwarf mass limit ($\simeq$ 13 $M_{J}$), supporting their status as planets. The lower limit of $v \sin i$ for HD 69830 and HD 92788 is approximately zero, which leads to a minimum inclination of 0 deg. As a result, the maximum planetary masses cannot be determined. 

The calculated mass of HD 92788 b (28 $M_{J}$) is found to be much larger than the brown dwarf limit due to the low inclination predicted from the rotation period ($8_{-8}^{+14}$ deg). The lower range of the mass (9 $M_{J}$) still allows the possibility that it is a planet, however there is evidence to suggest that it could be a very low-mass brown dwarf. Similarly, the inclination of HD 69830 is found to be low ($13_{-13}^{+27}$ deg) and as a result, the planets are calculated to be over four times the minimum mass. This suggests that the planets are in the Saturn-mass rather than Neptune-mass regime. 

HD 69830 and HD 92788 are the stars with the lowest $v \sin i$ which leads to the low inclinations. Due to the difficulty in measuring such small values, other determinations in the literature find higher values of $v \sin i$ (1.1 km s$^{-1}$ \citet{Lovis06} and 1.8 km s$^{-1}$ \citet{Mayor04}, respectively). Using these values pushes the inclinations to $58_{-32}^{+32}$ deg and $53_{-20}^{+37}$ deg, resulting in the planet masses being not much above the minimum mass. To determine the nature of these objects, a consensus on the value of $v \sin i$ must be found. 

HD 89744b has one of the highest minimum-mass companions discovered and may be a low mass brown dwarf if the inclination were low enough. The moderate inclination, $54_{-12}^{+36}$, leads to a mass range of 8--15 $M_{J}$ which does not rule this out; however, the calculated value falls below the brown dwarf limit, which suggests that the companion is a very massive planet. 

\section{CONCLUSIONS}

The \ion{Ca}{ii} H \& K records of 36 exoplanet host stars and photometric observations of HD 130322 were analysed for periodicities associated with stellar rotation. An increased \ion{Ca}{ii} H \& K dataset allowed previously unknown rotation periods to be determined. Ten stars exhibit periodicities that are consistent with $R_*$, $v \sin i$ and $P_{calc}$, the rotation period estimated from the average magnetic activity, and are therefore inferred to be due to rotational modulation. 

HD 3651, HD 130322 and HD 22049 show the recurrence of strong and similar periodicities over many seasons which we attribute to  rotation. The rotation period of HD 3651 has been updated from \citet{Bal96} to $40.2 \pm 4.0$ d and that of HD 22049 from \citet{Don96} to $11.3 \pm 1.1$d. We also present the rotation period of HD 130322, $P_{rot}=26.1 \pm 3.5$ d, which has not previously been reported. 

The seven other stars show periodicities in only one or two seasons so the rotation period is less firmly determined (HD 9826, HD 10697, HD 6930, HD 89744, HD 92788, HD 154345 and HD 217014). Four of these stars' rotation periods have not previously been reported (HD10697, HD 69830, HD 92788 and HD 154345), while the grade assigned to the rotation period of HD 217014 has been raised from \textit{weak} to \textit{probable} \citep{Henry00}. Further monitoring and analysis of the \ion{Ca}{ii} H \& K or photometric fluxes of these stars could increase the reliability of the estimated rotation periods. None of the rotation periods refute the inference of an orbiting companion in favour of rotational modulation.

The inclination of the stellar rotation axes was calculated, however the planets presented here are non-transiting, therefore other methods must be used to determine their orbital inclination. It is hoped that future astrometric observations will be able to determine and compare $i_{p}$ with the inclination of the stellar rotation axes reported here and to infer the line-of-sight alignment of these systems in order to constrain theories of planet formation and evolution.

Planetary masses were inferred under the assumption that the stellar rotation and planetary orbital axes are aligned. This method allowed the maximum mass of the orbiting companions to be constrained for most of the systems. All but one of the companions have a calculated mass below the brown dwarf limit, thereby supporting the inference of their planetary nature. The calculated mass of HD 92788 b, 28 $M_{J}$, suggests that it may be a low-mass brown dwarf and warrants further investigation. 

\bibliography{bib_old}

    \begin{figure*}
    
 \appendix
 \section{Light curves and periodograms showing seasons containing rotation modulation in ten exoplanet host stars.}   
The figures have the following features: Vertical dashed lines in the periodograms to indicate the orbital period of the planet(s) (the longer orbital periods do not appear on the plot as the orbital frequency is close to zero on the scale); dotted horizontal lines showing the cut-off for a significant detection of a period, FAP = 0.1. Occasionally there is a second significant peak in the periodogram. In the majority of cases, this disappeared when the primary peak was removed and is therefore likely to be a harmonic of the main peak. Those cases where this does not occur are mentioned in the text.\\[3ex]
 
    	\caption{Seasons showing rotational periodicities in HD 3651. The peaks are strong and so the periodicity is graded as \textit{confirmed}.}
	\centering
    	\begin{tabular}{cc}
     		\includegraphics[angle=0, width=80mm]{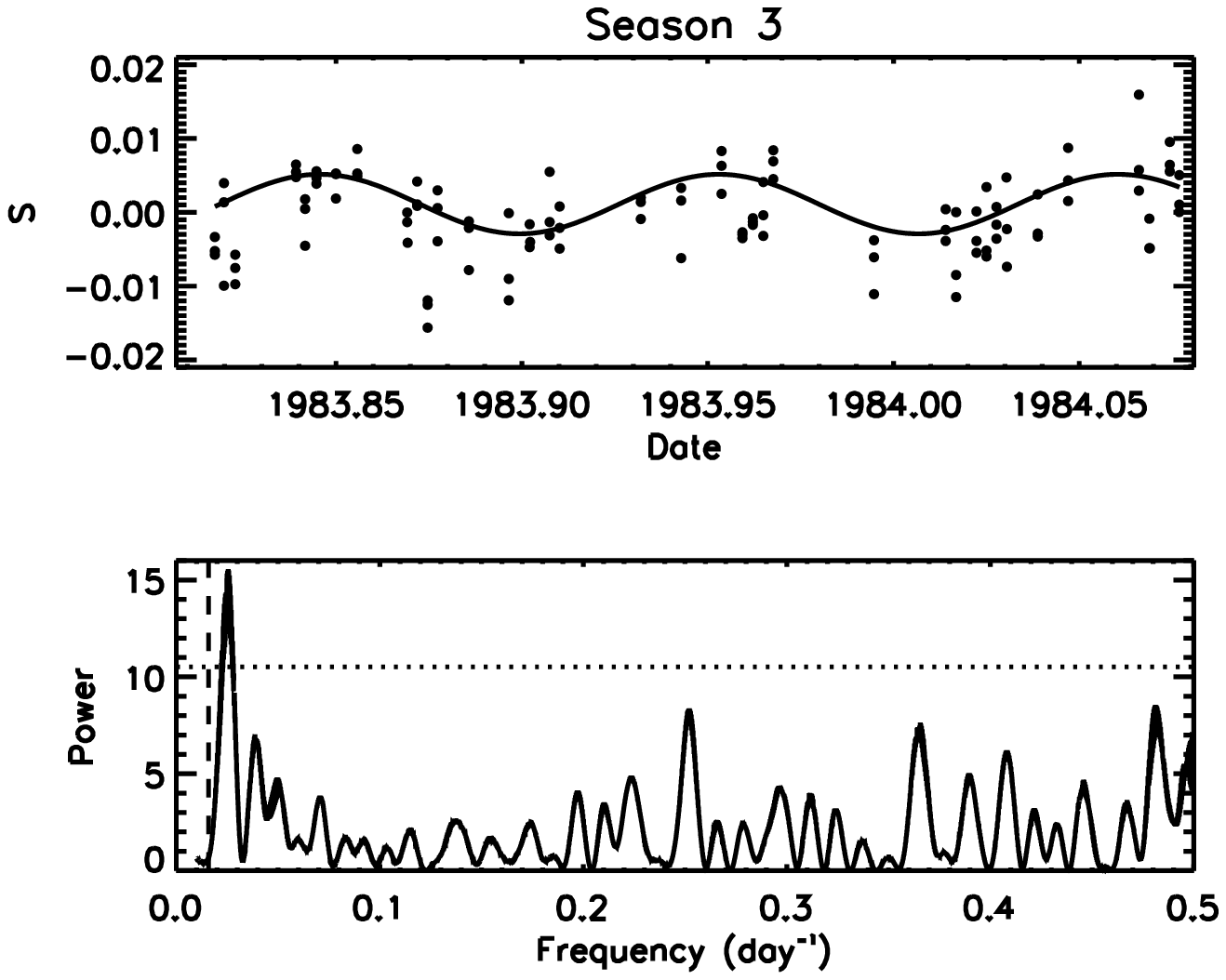} &
		\includegraphics[angle=0, width=80mm]{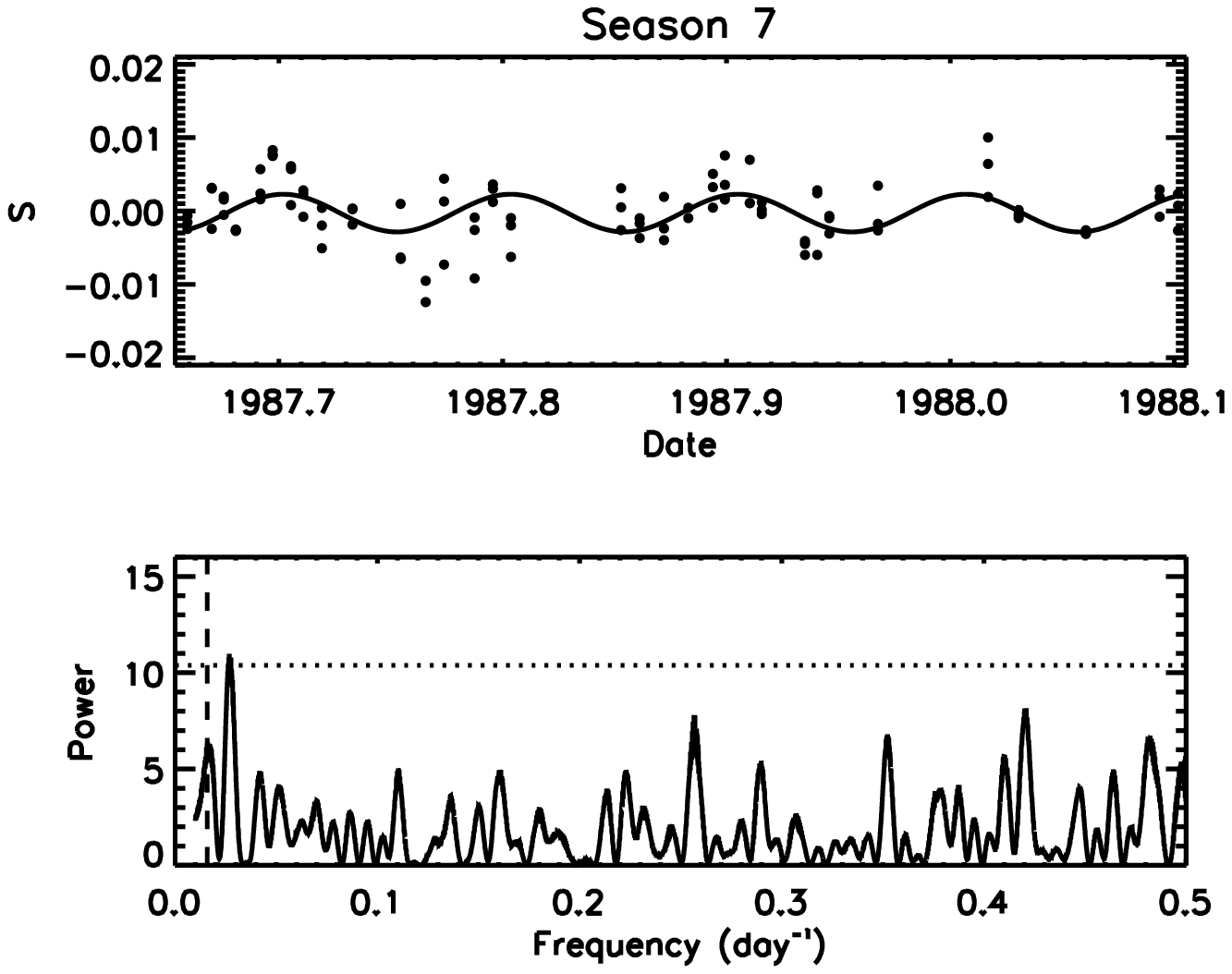}\\[3ex]
		\includegraphics[angle=0, width=80mm]{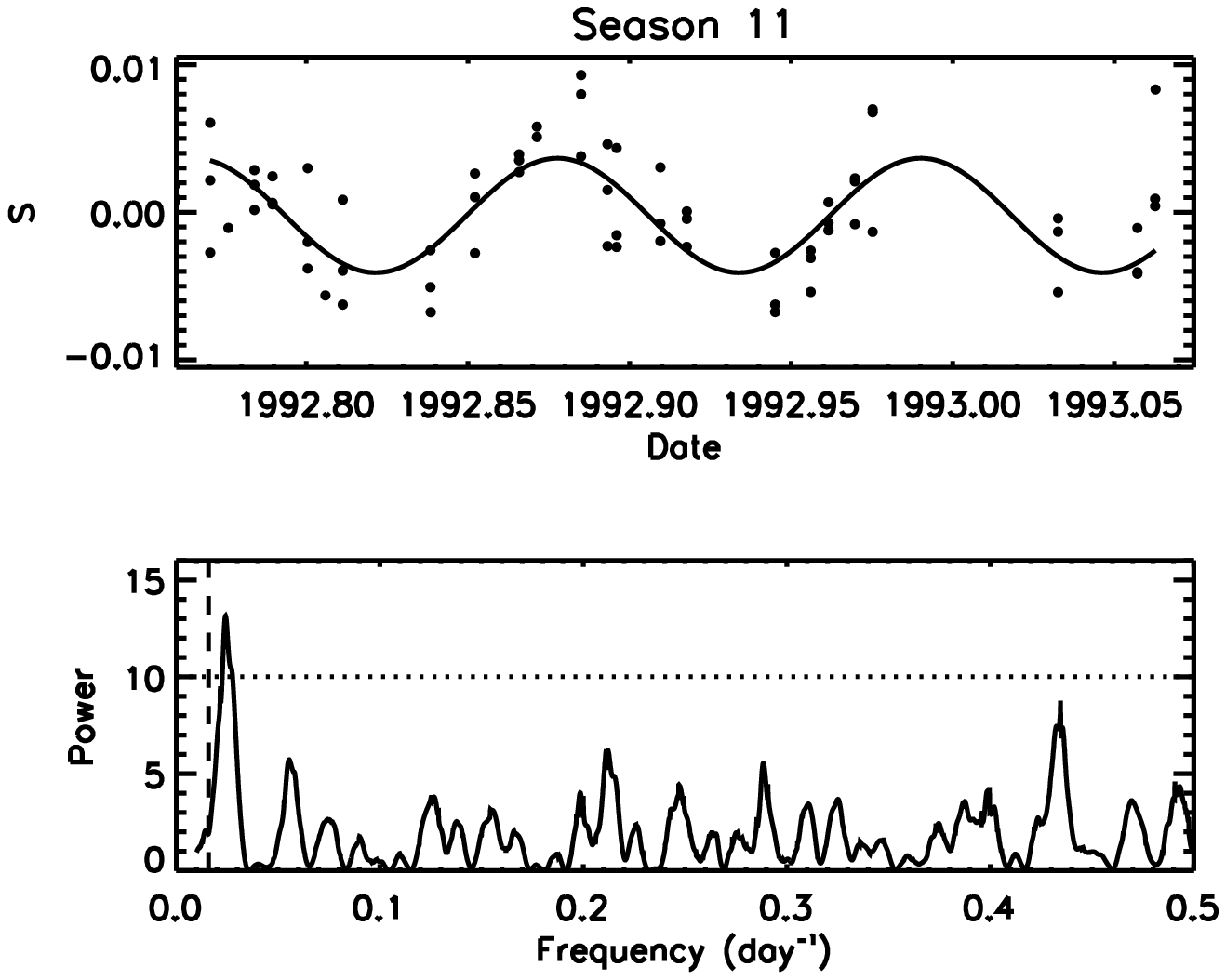}&
		\includegraphics[angle=0, width=80mm]{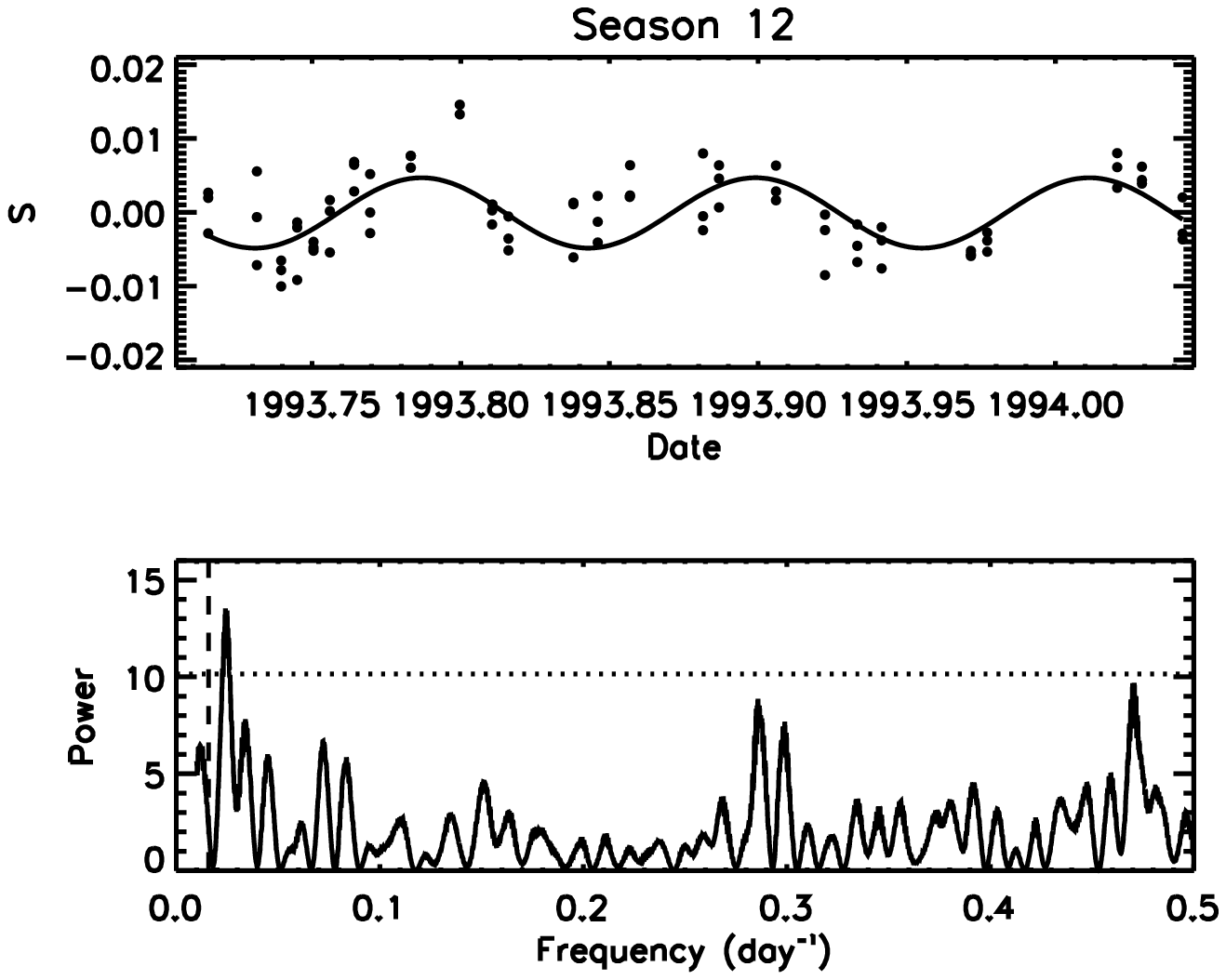}\\[3ex]
		\includegraphics[angle=0, width=80mm]{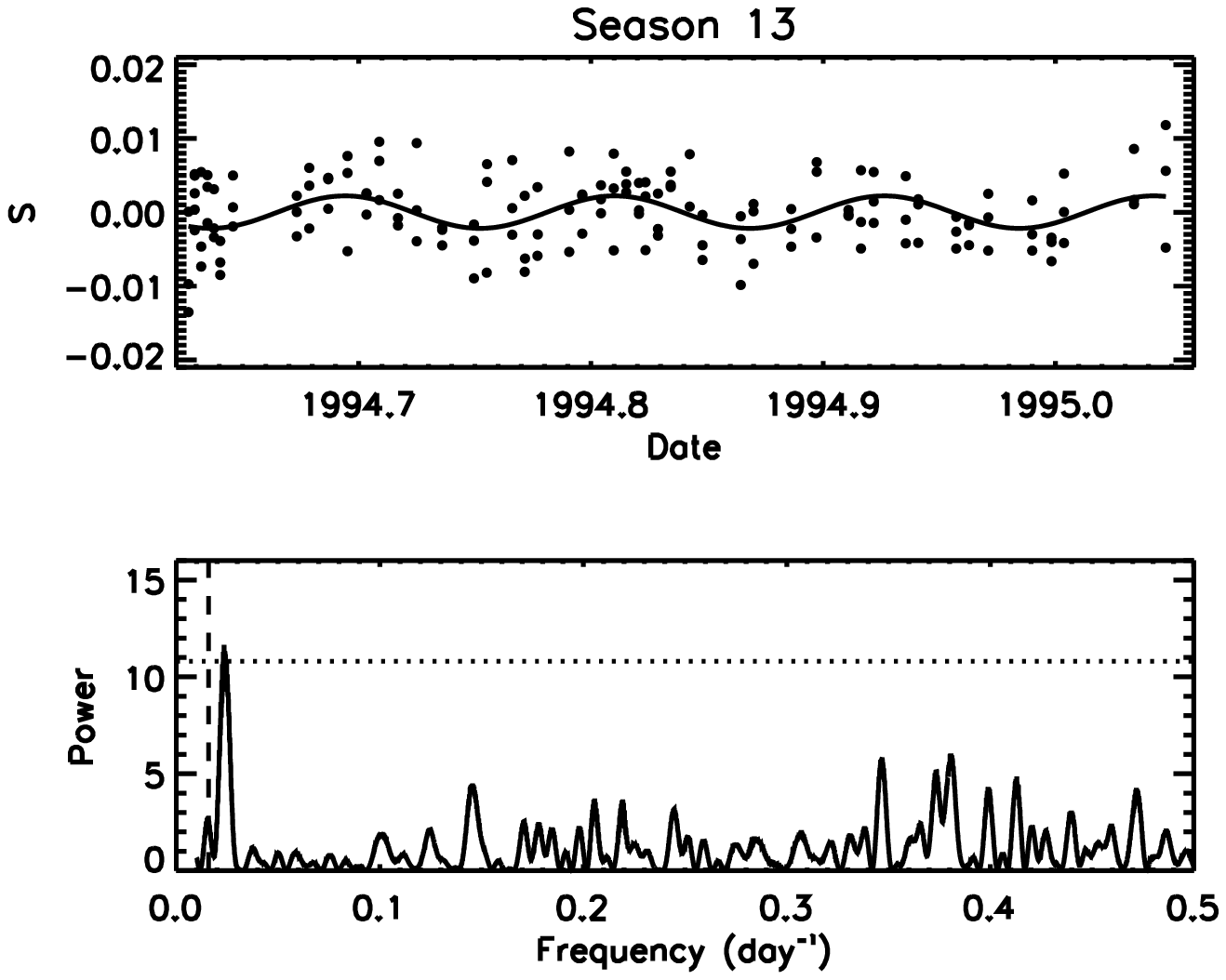} &
		\includegraphics[angle=0, width=80mm]{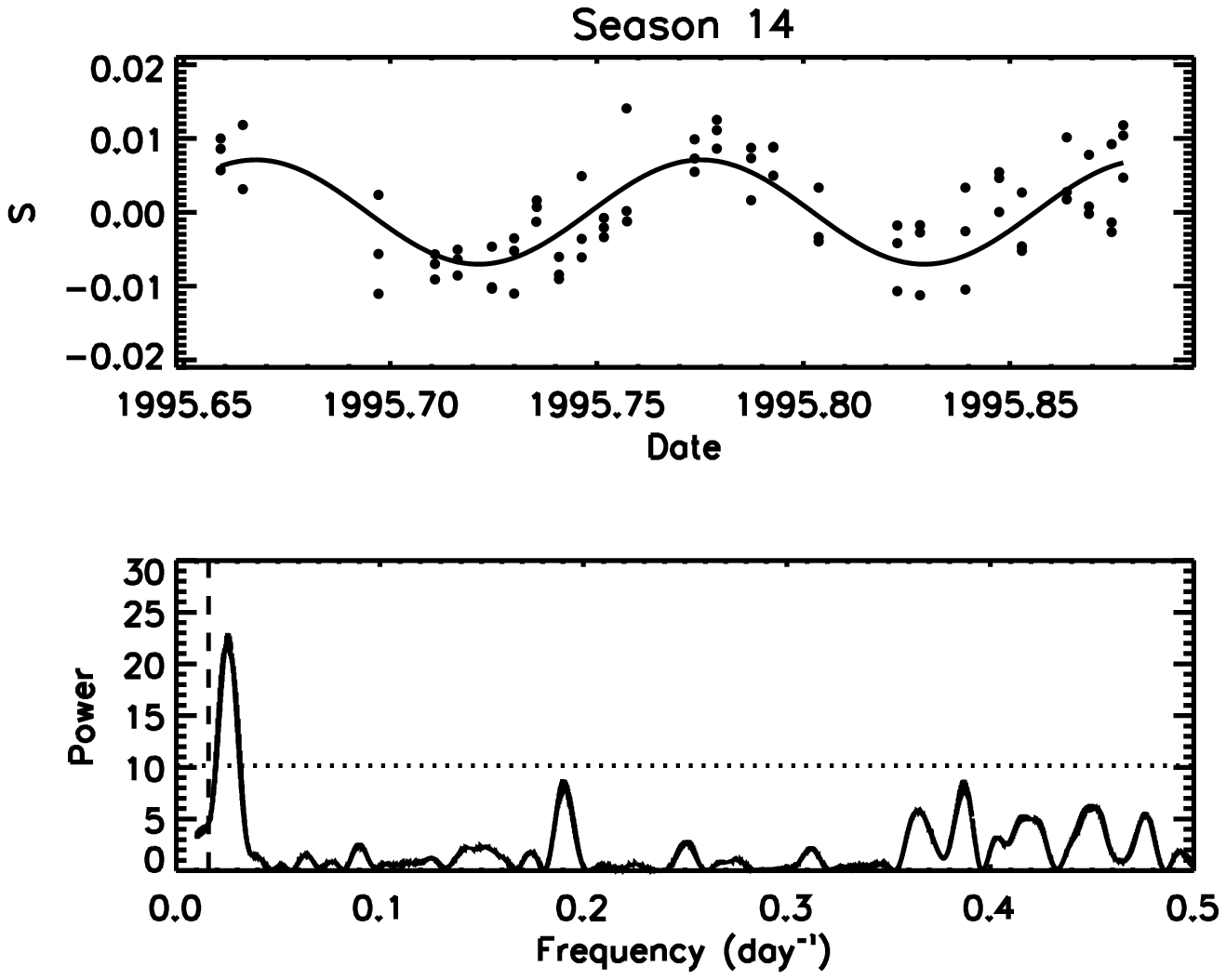} \\
   	 \end{tabular}
   \end{figure*}
     
   \begin{figure*}
   	\begin{tabular}{c}
		\includegraphics[angle=0, width=80mm]{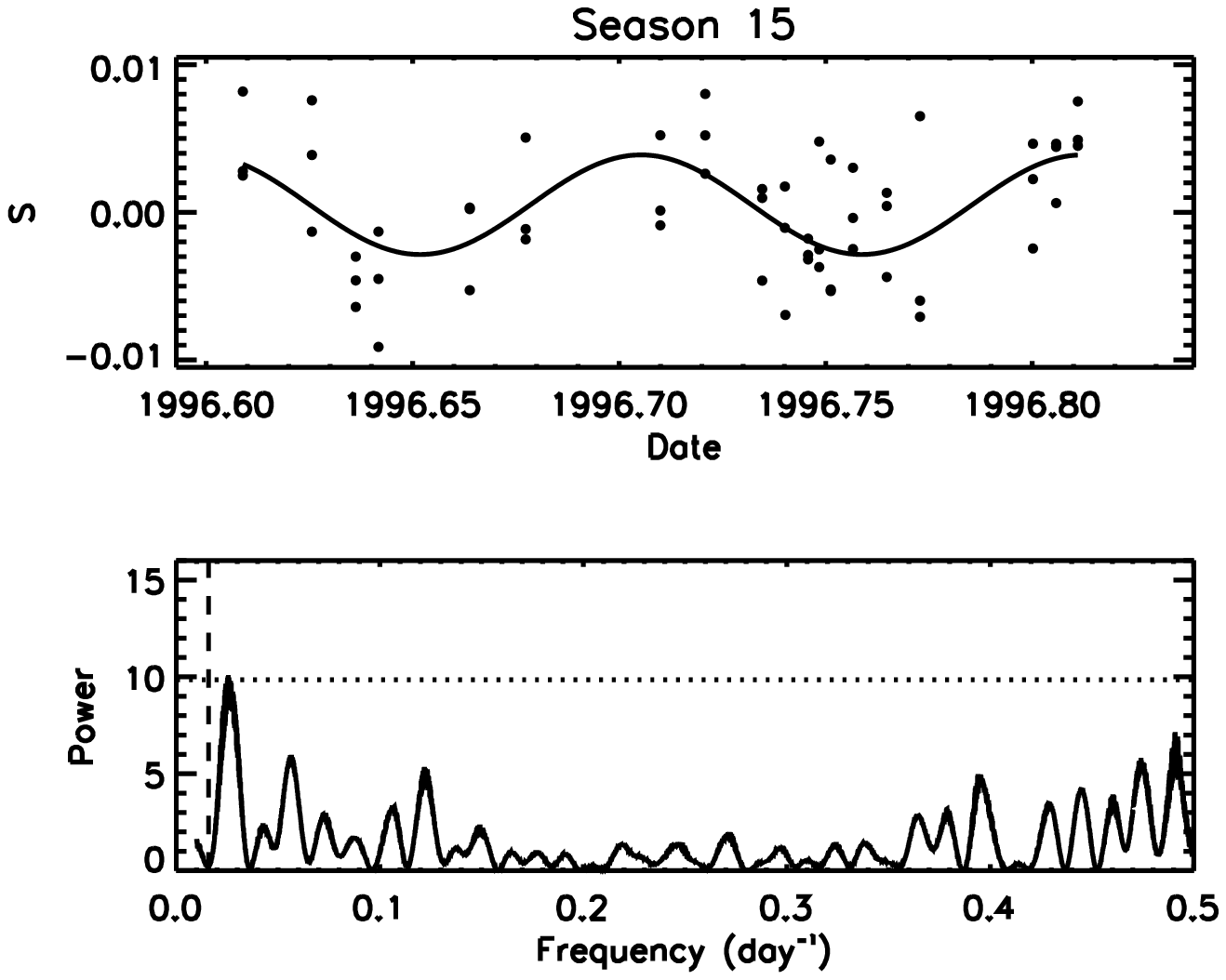}\\ [3ex]
	 \end{tabular}
   \end{figure*}

    \begin{figure*}
    	\caption{Only one of six seasons shows a period consistent with rotation in HD 9826 ($\upsilon$ And). It is therefore graded as \textit{weak}. }
	\centering
    	\begin{tabular}{c}
     		\includegraphics[angle=0, width=80mm]{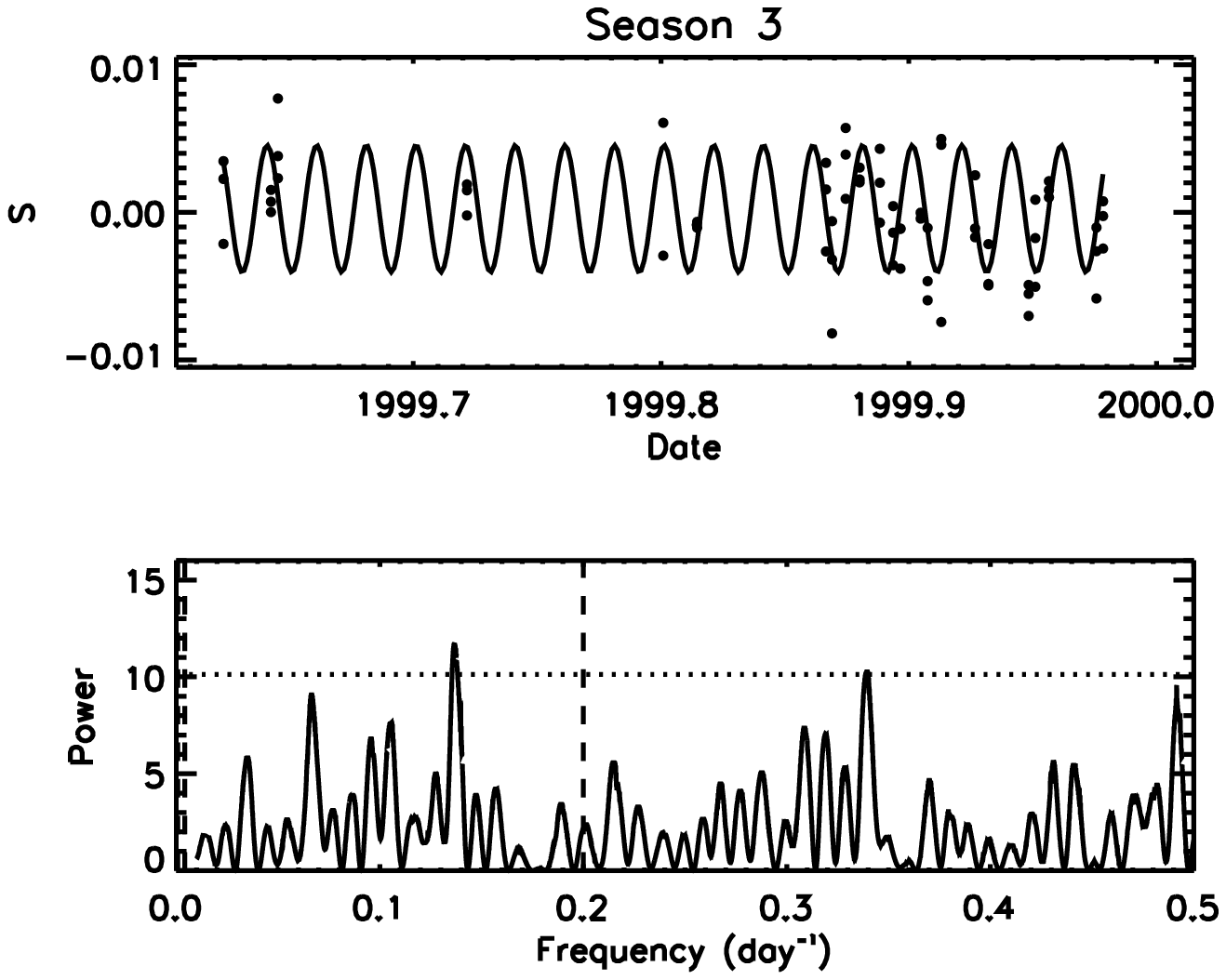} \\[3ex]
   	 \end{tabular}
   \end{figure*}
     
    \begin{figure*}
    	\caption{One season in HD 10697 shows a steep trend (left figure). When the trend is removed (right figure), a period of $\sim$ 33 d is revealed, which may be due to rotation. It is not repeated in any other season, so is graded as \textit{weak}.}
	\centering
    	\begin{tabular}{cc}
     		 \includegraphics[angle=0, width=80mm]{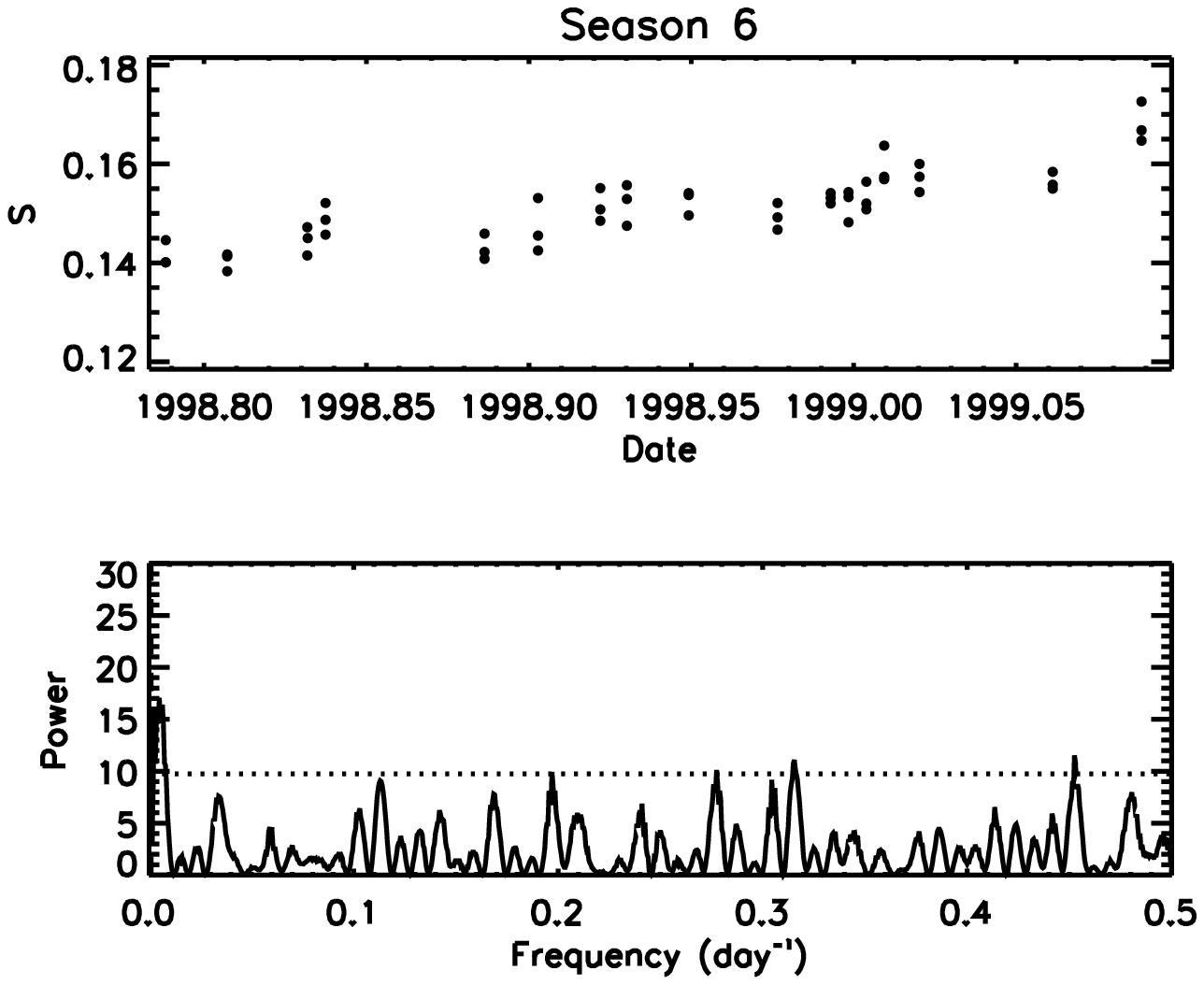}&
		  \includegraphics[angle=0, width=80mm]{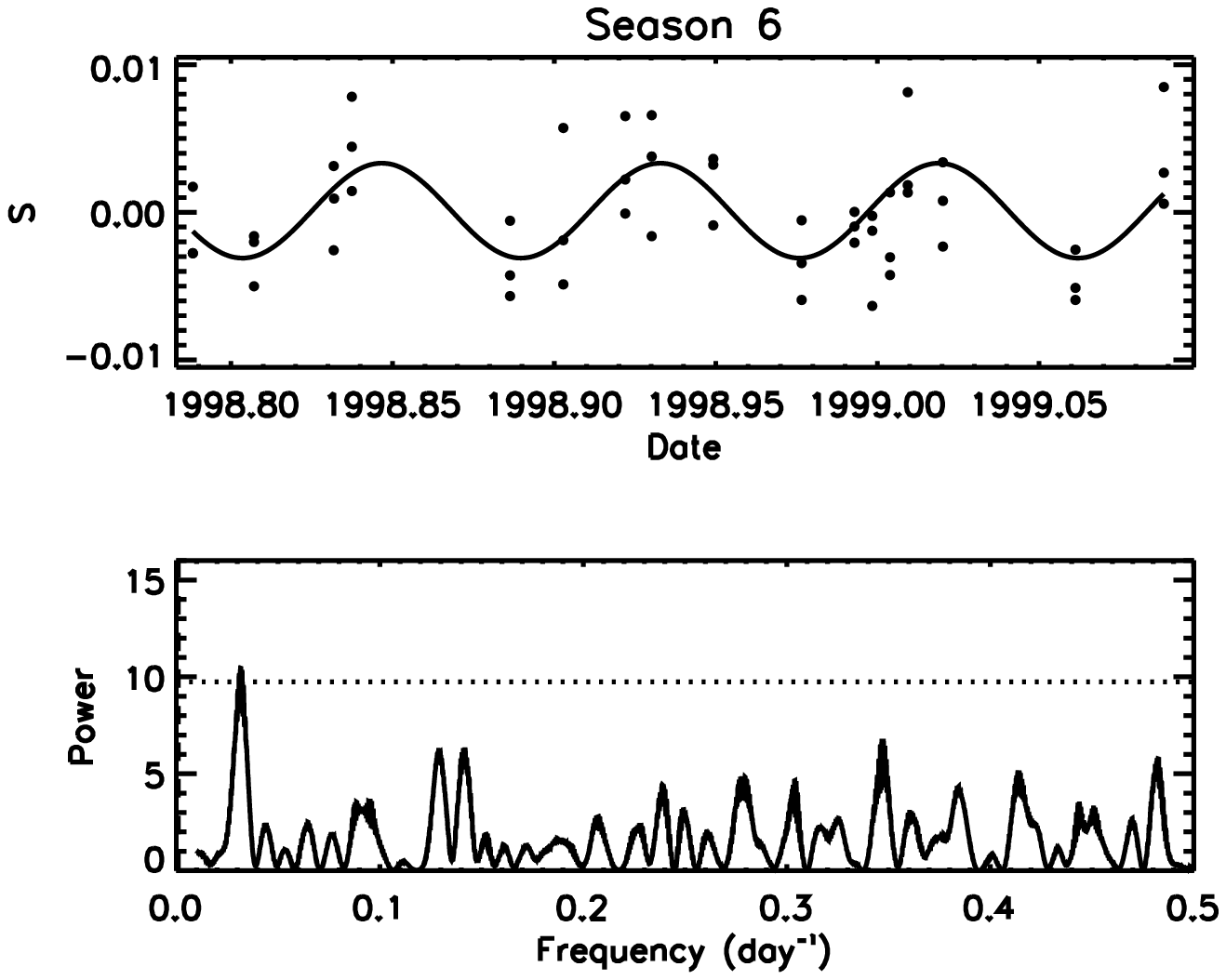} \\
   	 \end{tabular}
   \end{figure*}

    \begin{figure*}
    	\caption{The rotational modulation in HD 22049 ($\epsilon$ Eri) is strong and persistent in nine seasons. It is graded as \textit{confirmed}.  }
	\centering
    	\begin{tabular}{cc}
     		\includegraphics[angle=0, width=80mm]{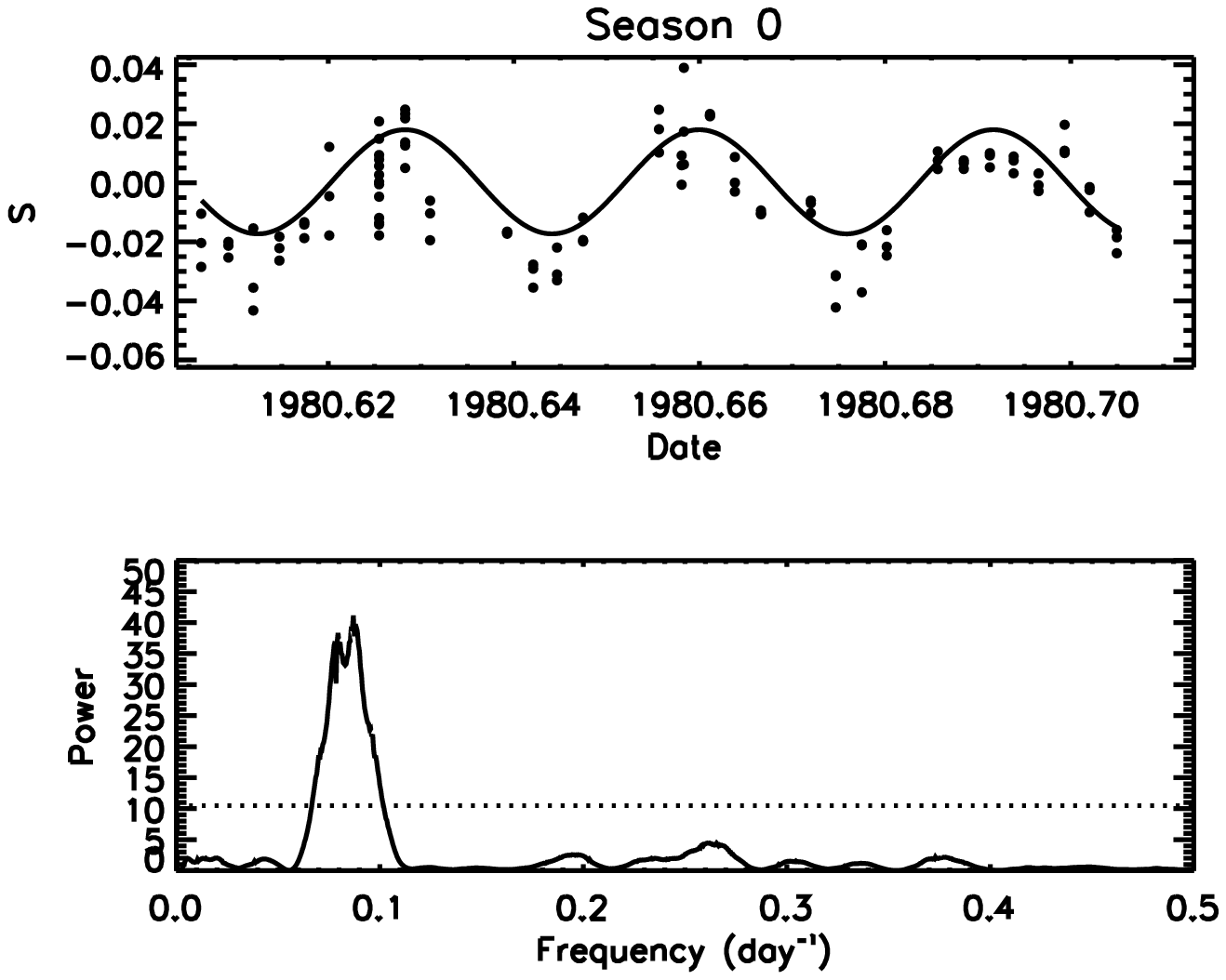} &
		\includegraphics[angle=0, width=80mm]{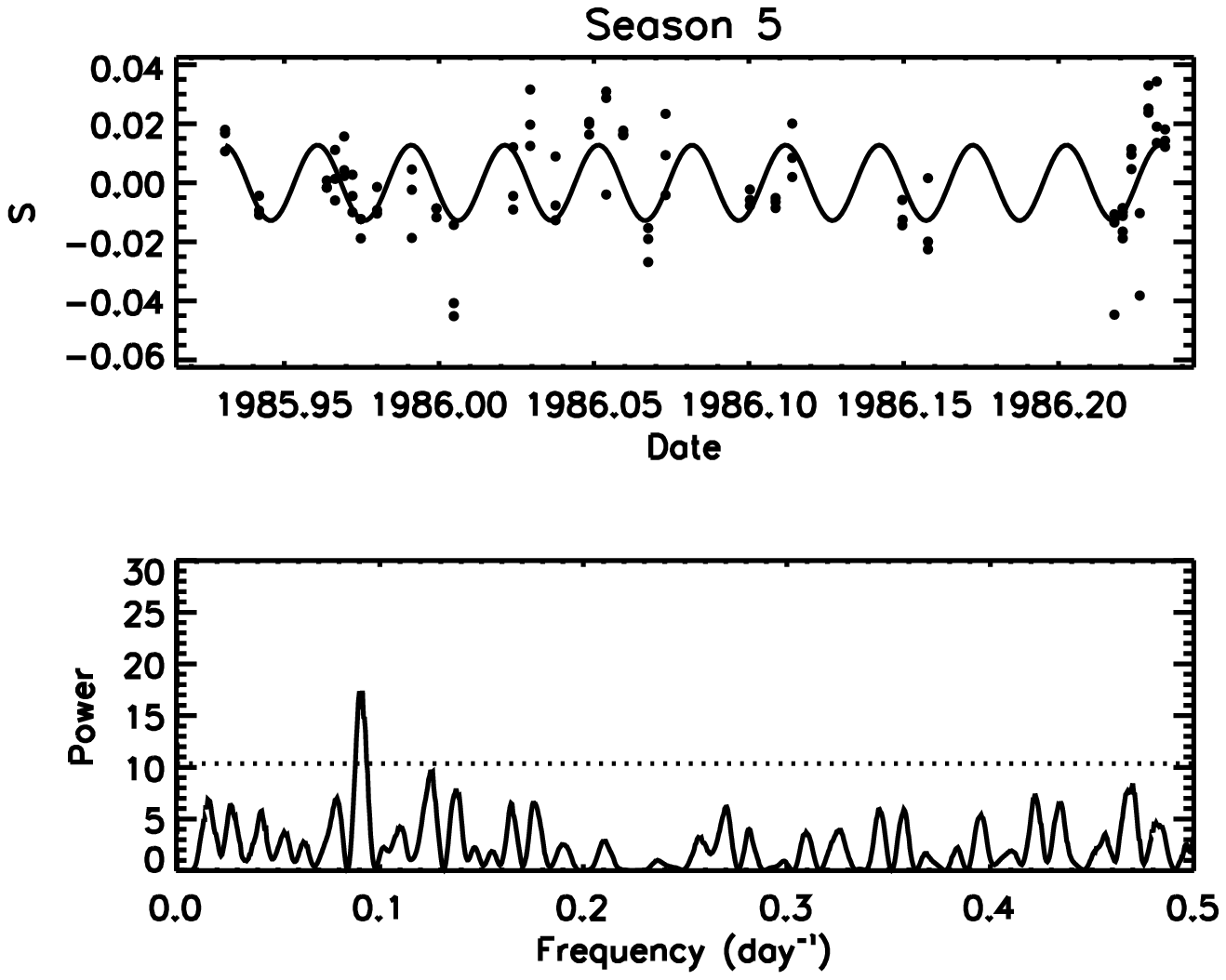}\\[10ex]
		\includegraphics[angle=0, width=80mm]{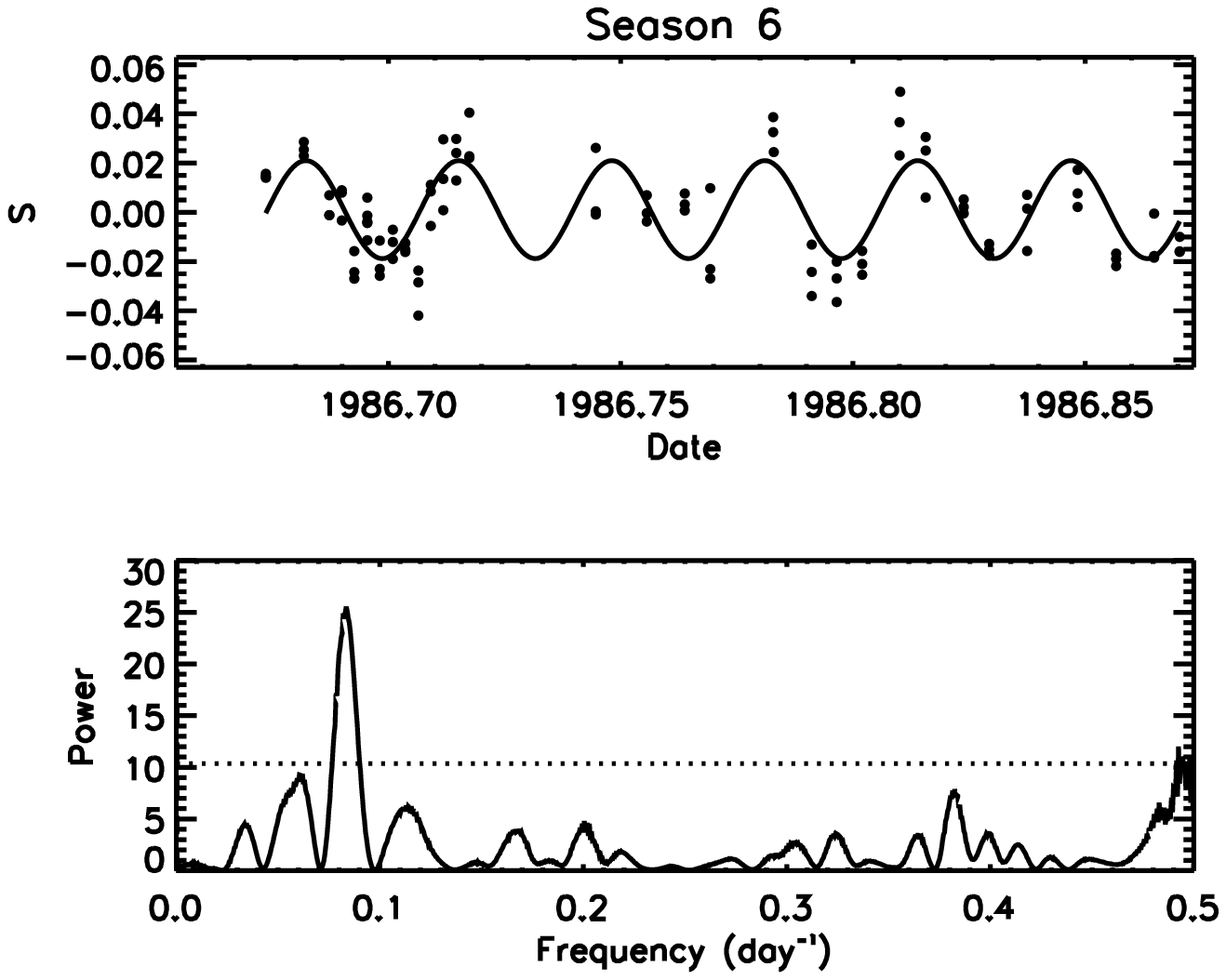}&
		\includegraphics[angle=0, width=80mm]{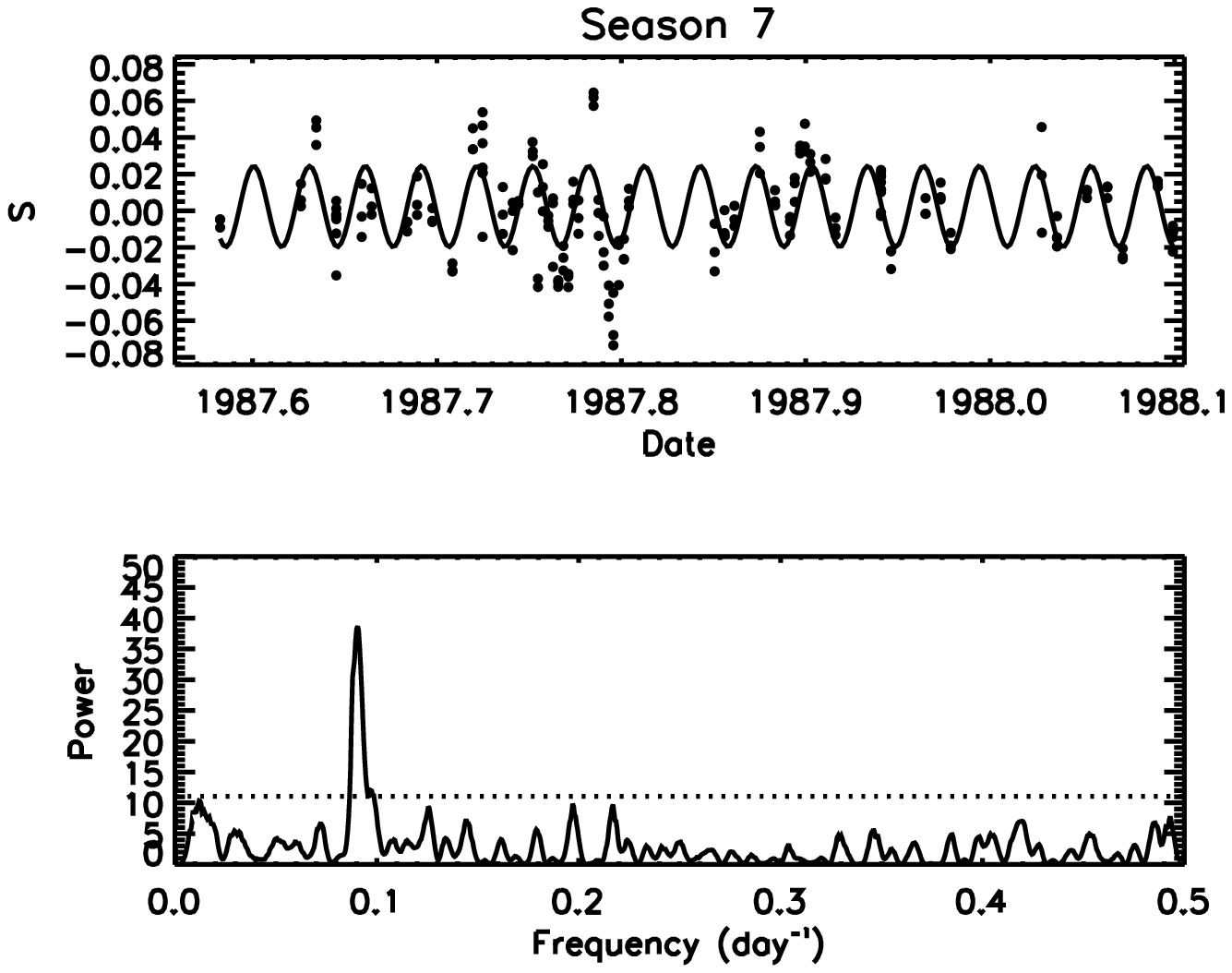}\\[10ex]
		\includegraphics[angle=0, width=80mm]{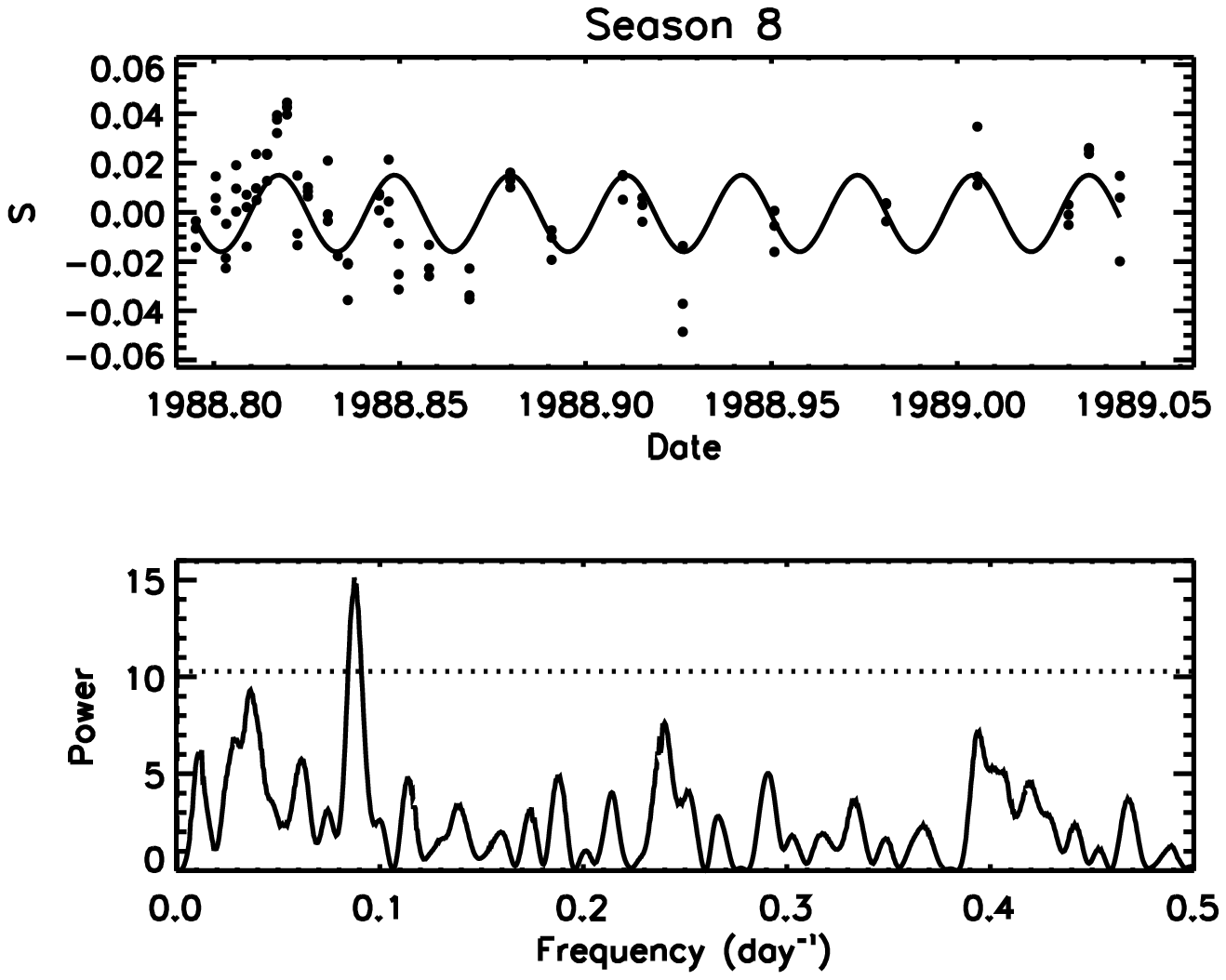} &
		\includegraphics[angle=0, width=80mm]{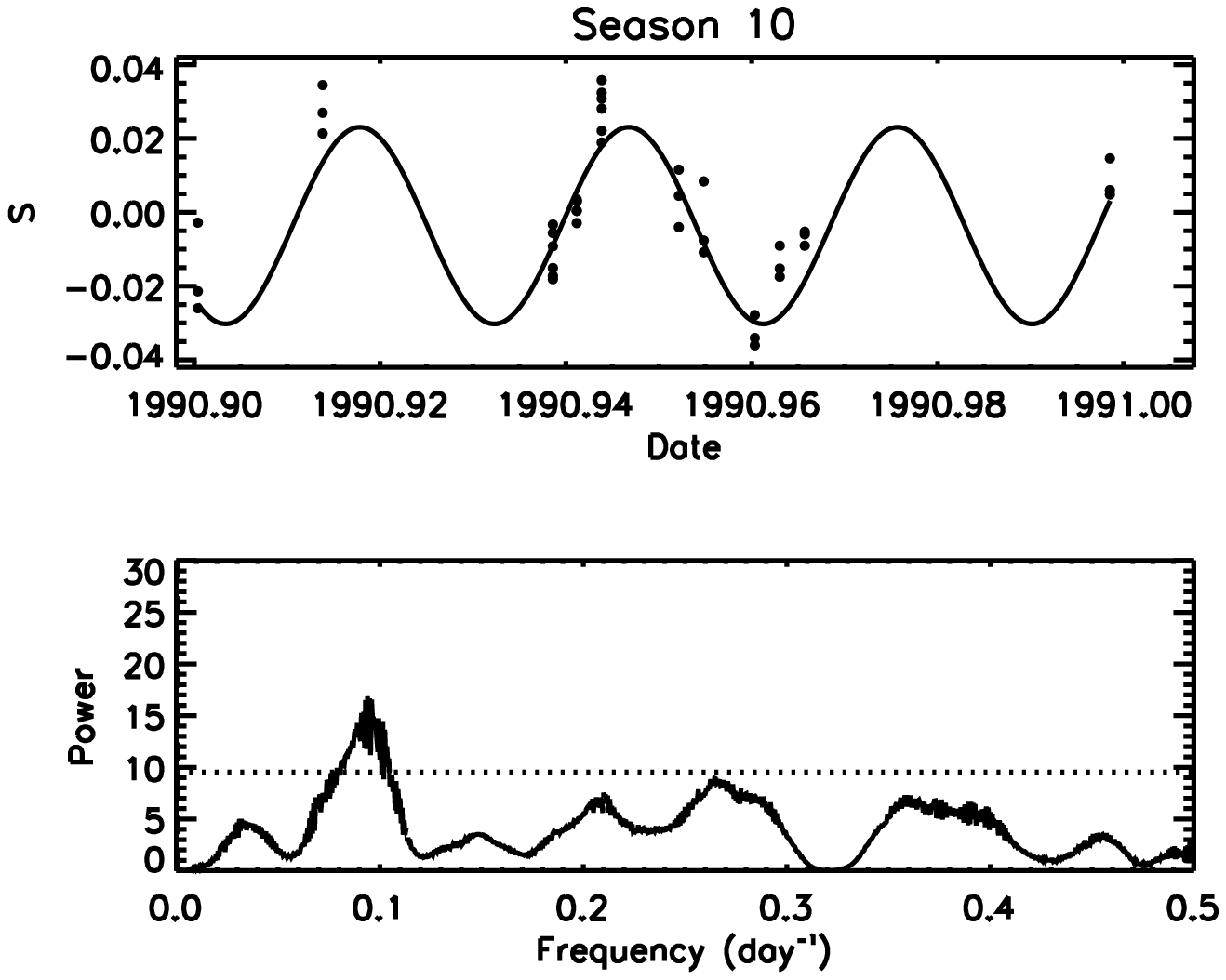}\\	
   	 \end{tabular}
   \end{figure*}
   
   \begin{figure*}
   \begin{tabular}{cc}
   
   		\includegraphics[angle=0, width=80mm]{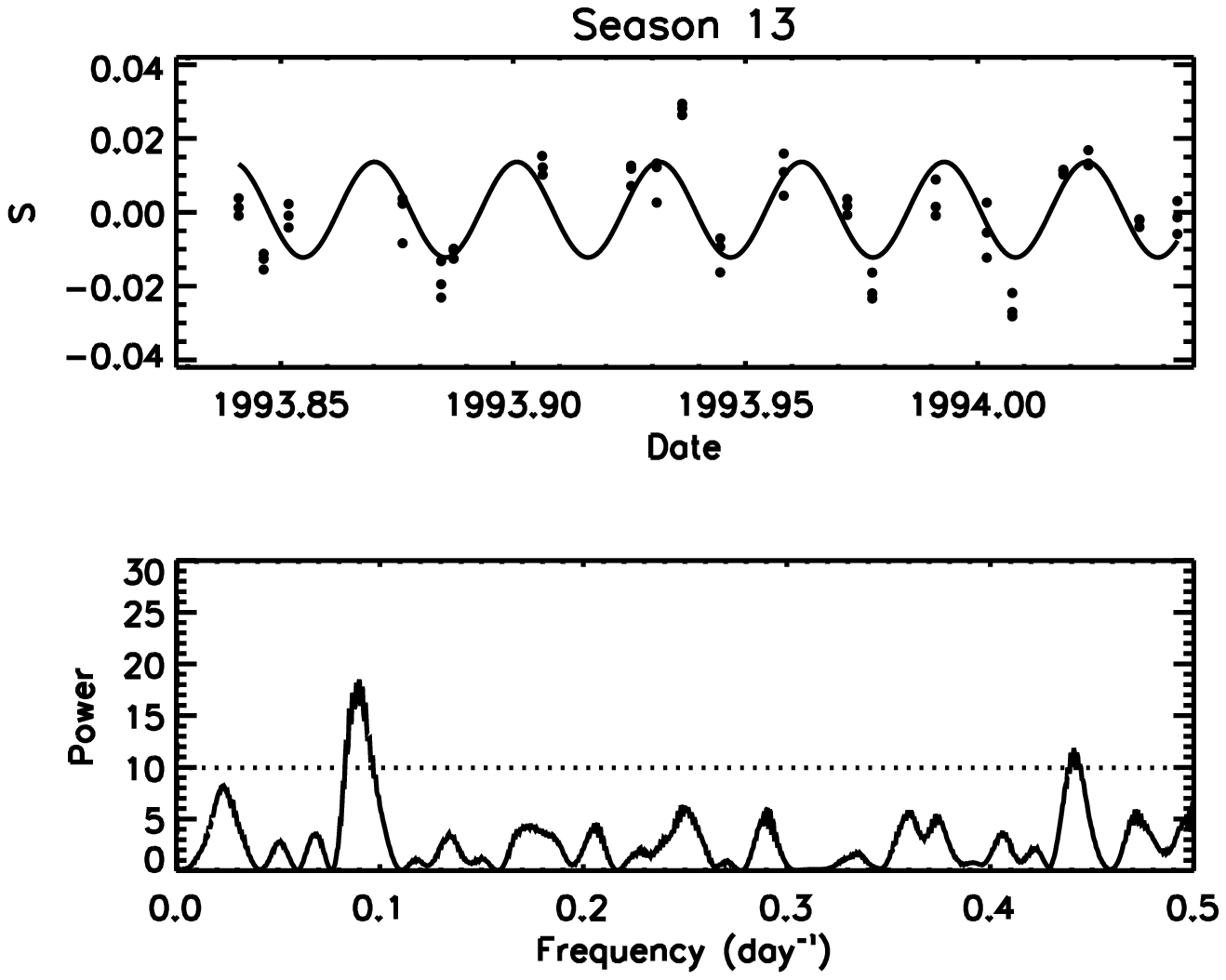} &
		\includegraphics[angle=0, width=80mm]{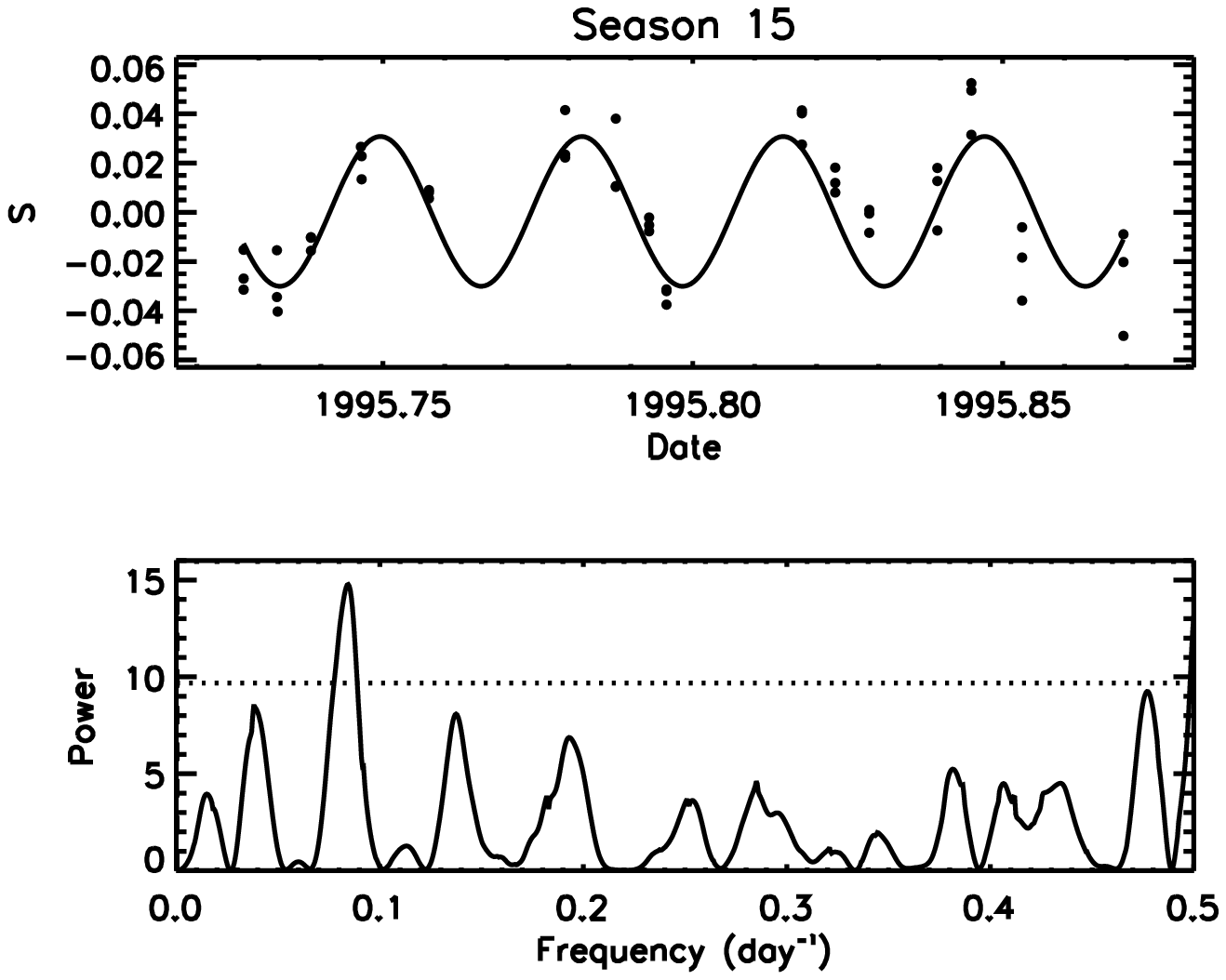} \\[10ex]
		\multicolumn{2}{c}{\includegraphics[angle=0, width=80mm]{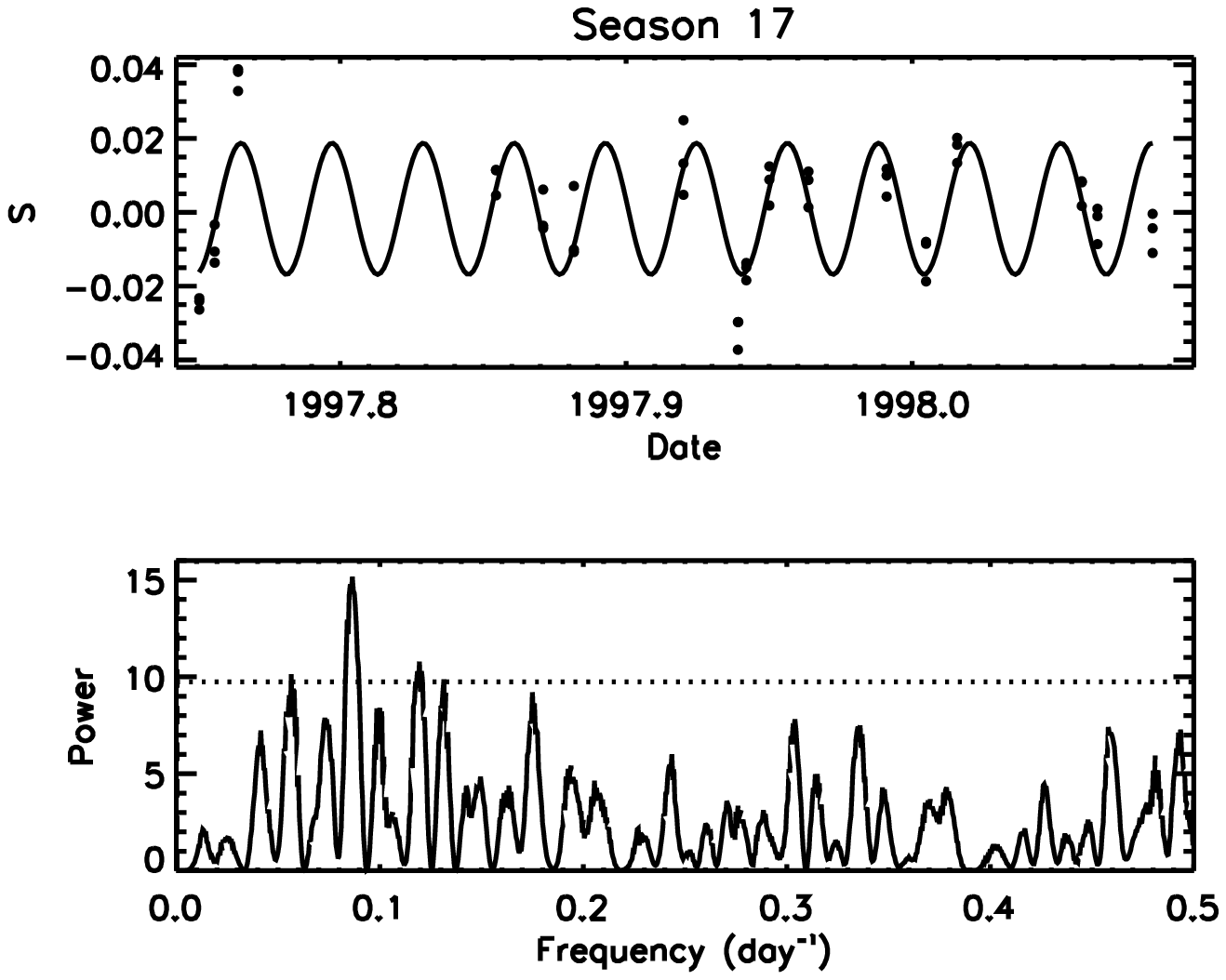} }\\[10ex]
	\end{tabular}
   \end{figure*}
     
    \begin{figure*}
    	\caption{A strong peak is visible in one season of HD 69830, however it is not repeated in another season so it is graded as \textit{probable}. The rotation period (35 d) is close to but not exactly the same as the orbital period of the second planet (32 d, dashed vertical line), see \ref{activity} for details. }
	\centering
    	\begin{tabular}{c}
     	\includegraphics[angle=0, width=80mm]{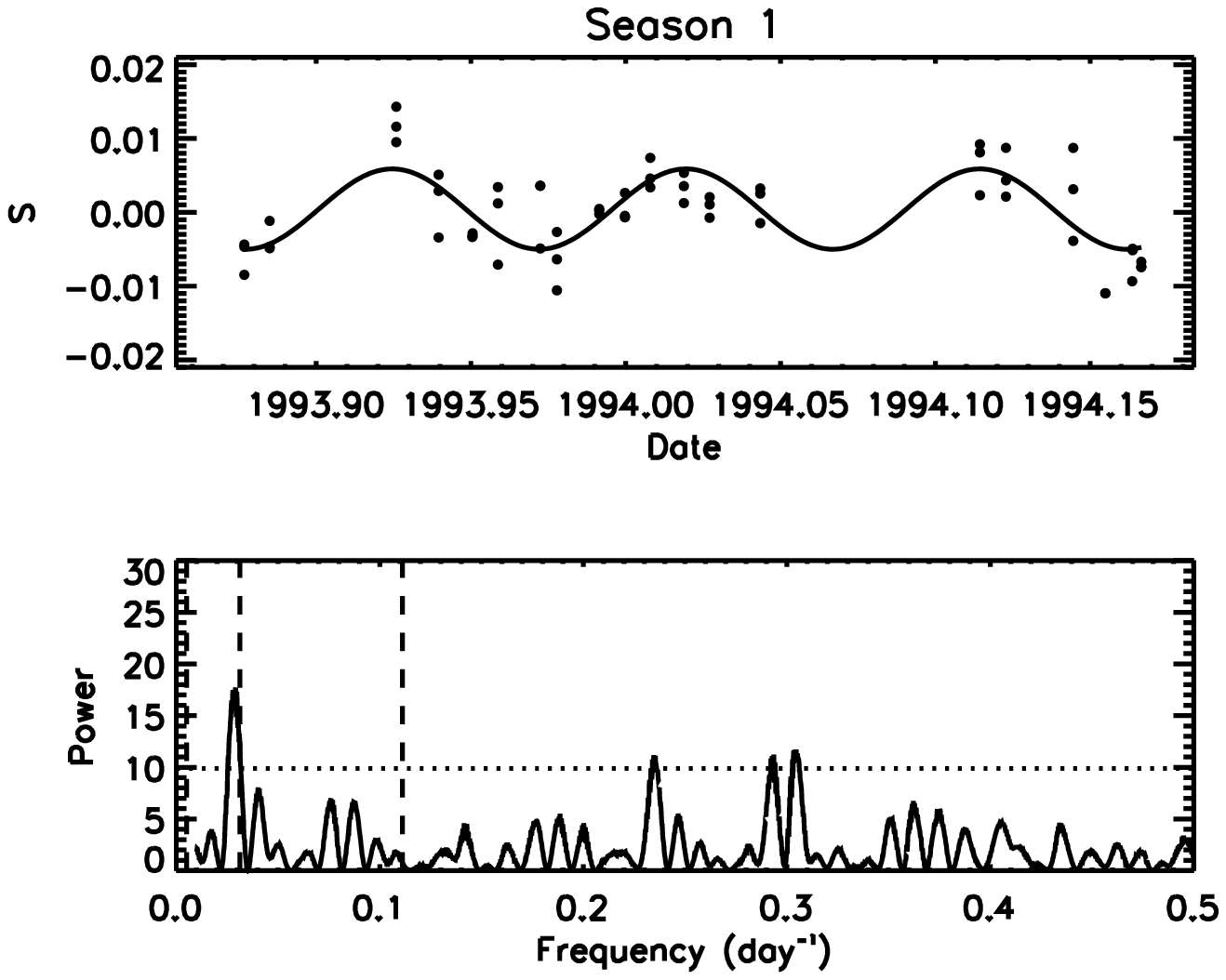} \\
   	 \end{tabular}
   \end{figure*}

    \begin{figure*}
    	\caption{Two distinct periods appear in HD 89744; 9 d and 12 d. The 9 d period is consistent with $R_{*}$ and $v \sin i$ and is designated a grade of \textit{probable}.}
	\centering
    	\begin{tabular}{cc}
     		\includegraphics[angle=0, width=80mm]{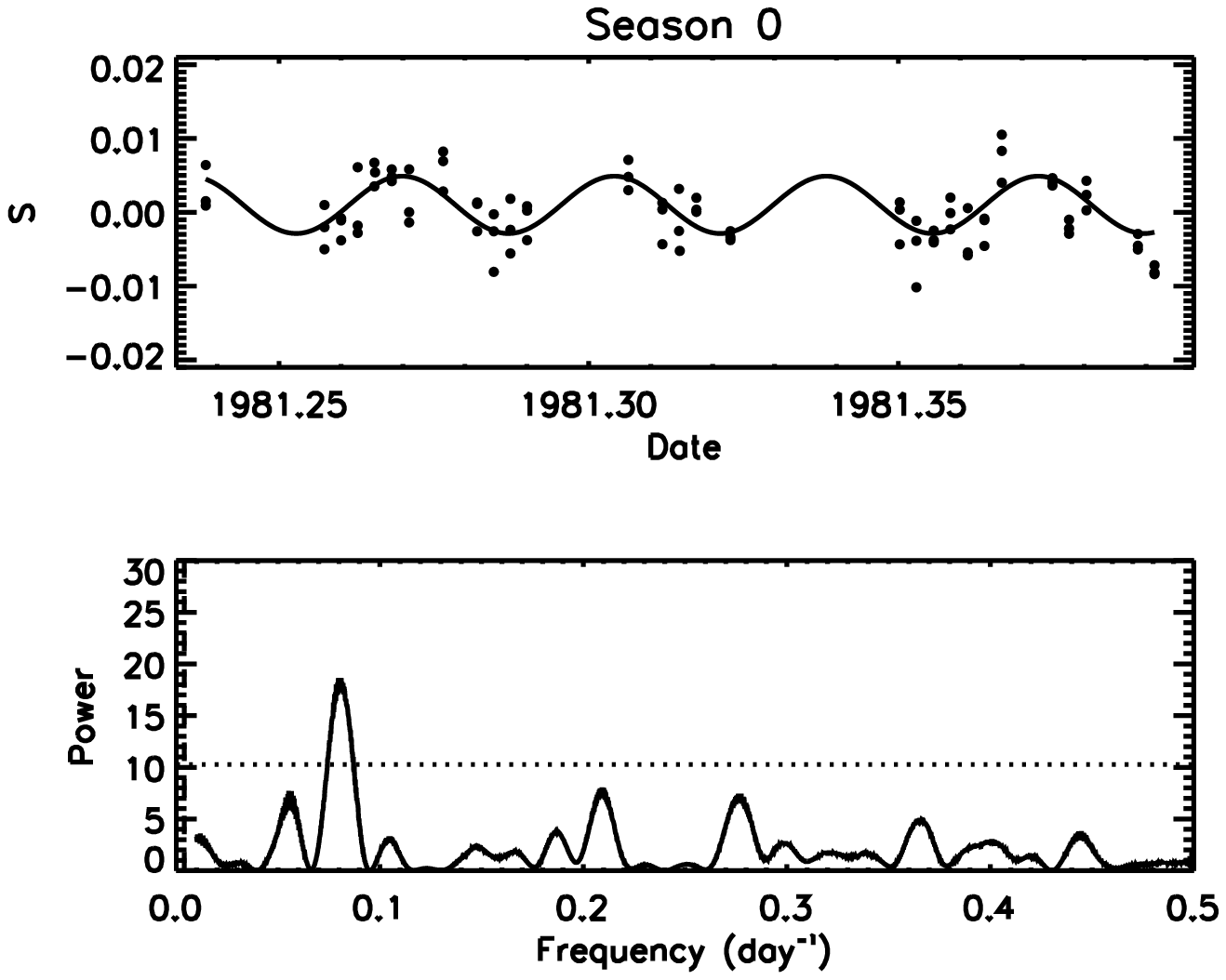} &
		\includegraphics[angle=0, width=80mm]{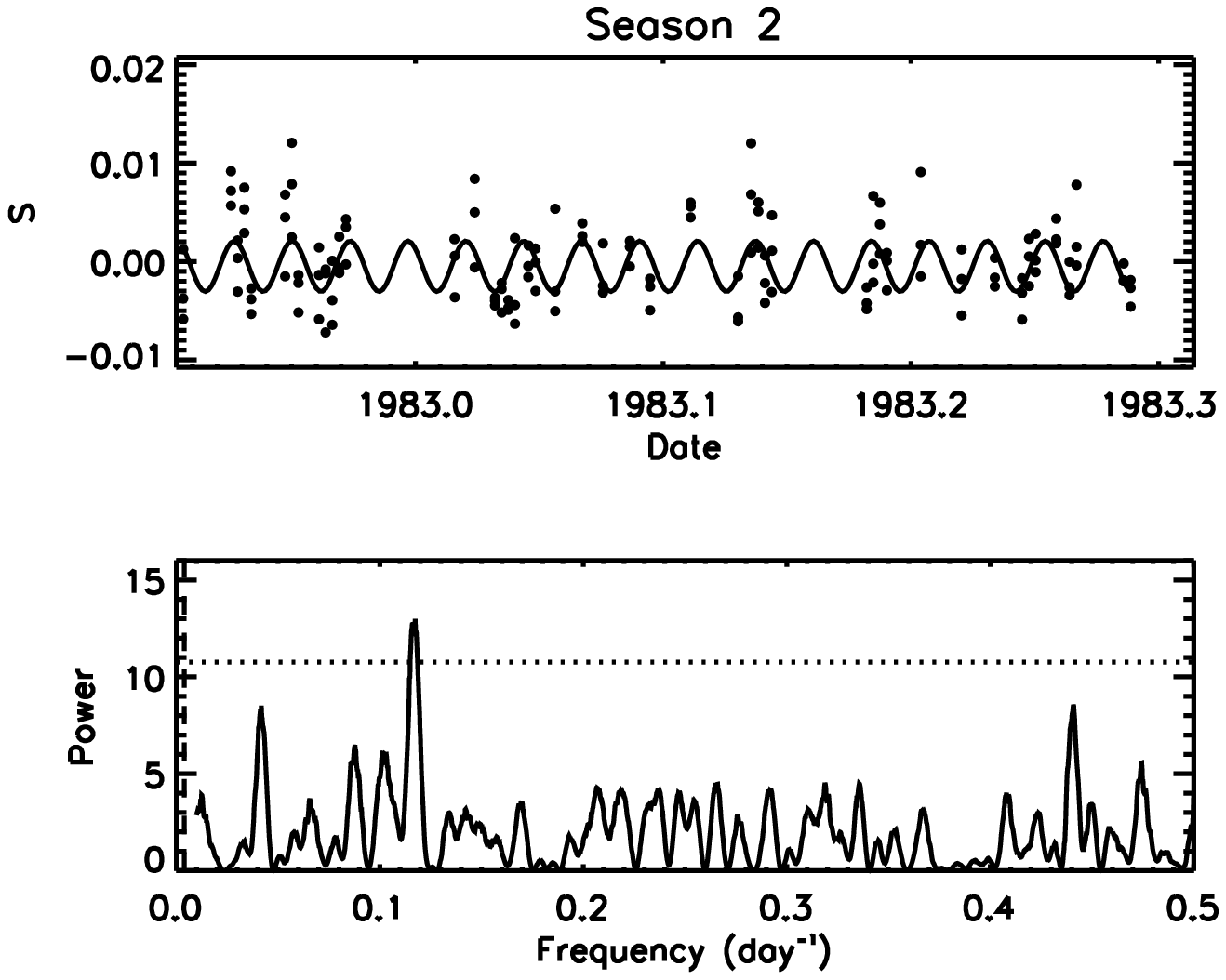}\\[10ex]
		\includegraphics[angle=0, width=80mm]{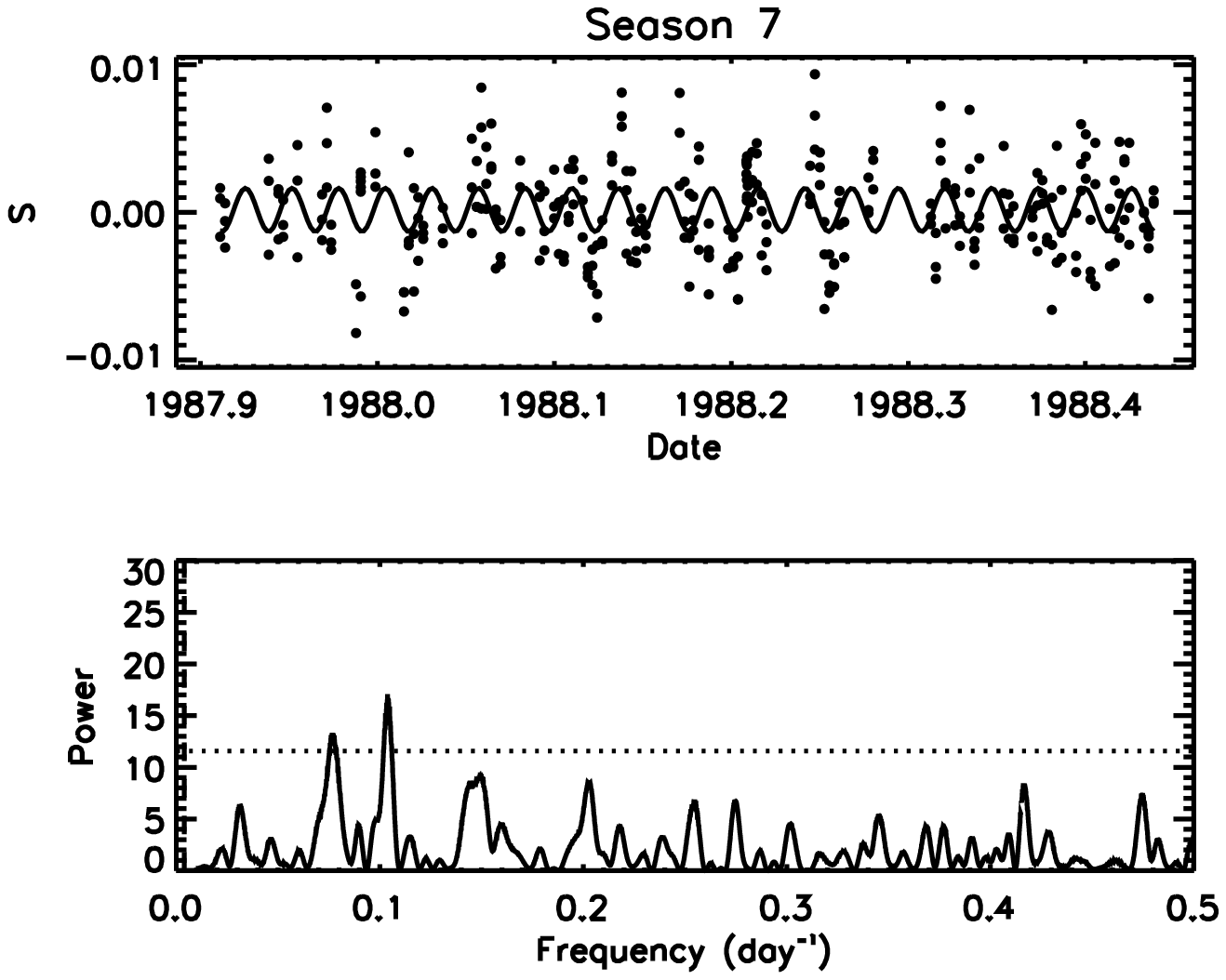} &
		\includegraphics[angle=0, width=80mm]{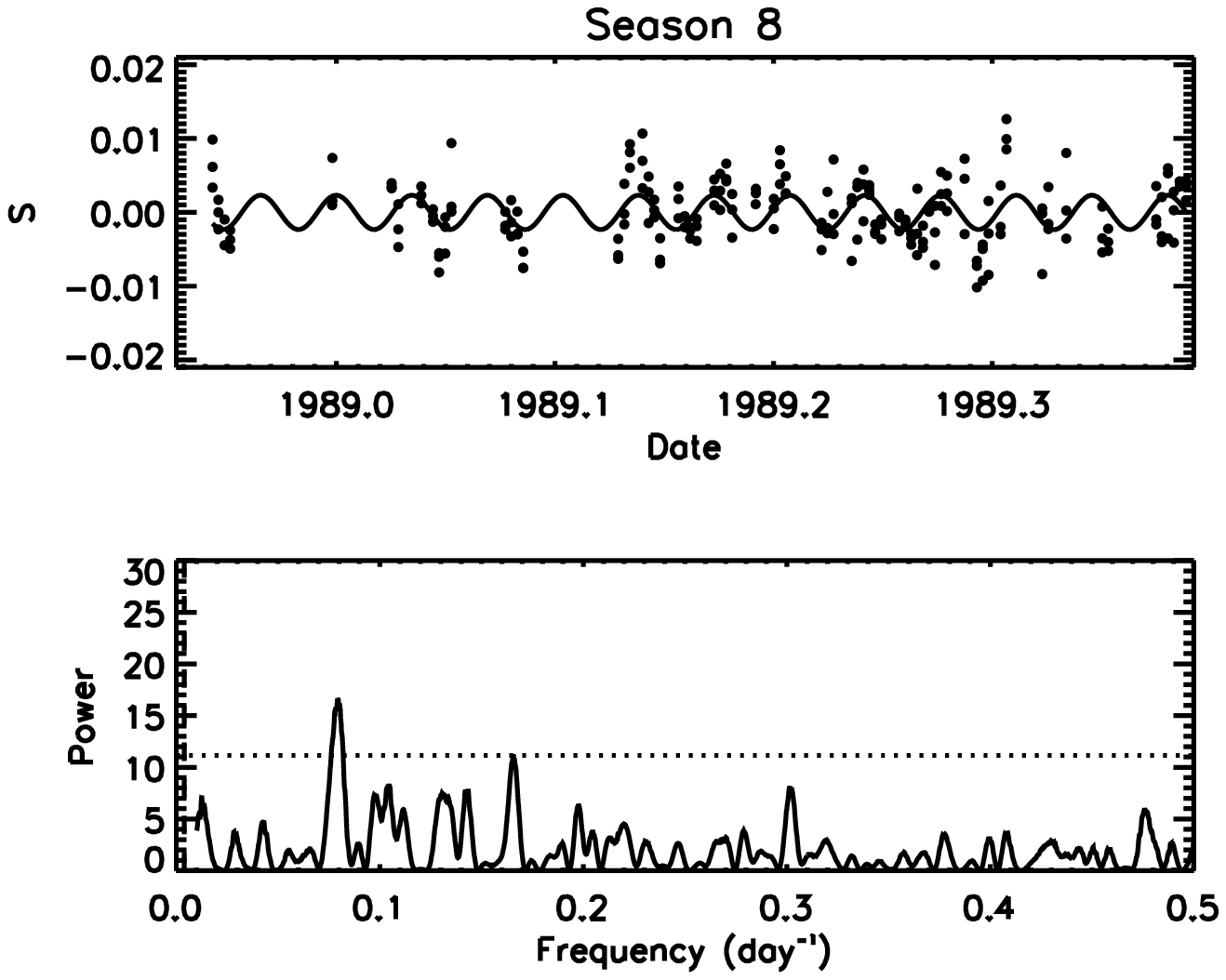} \\ [10ex]
		\includegraphics[angle=0, width=80mm]{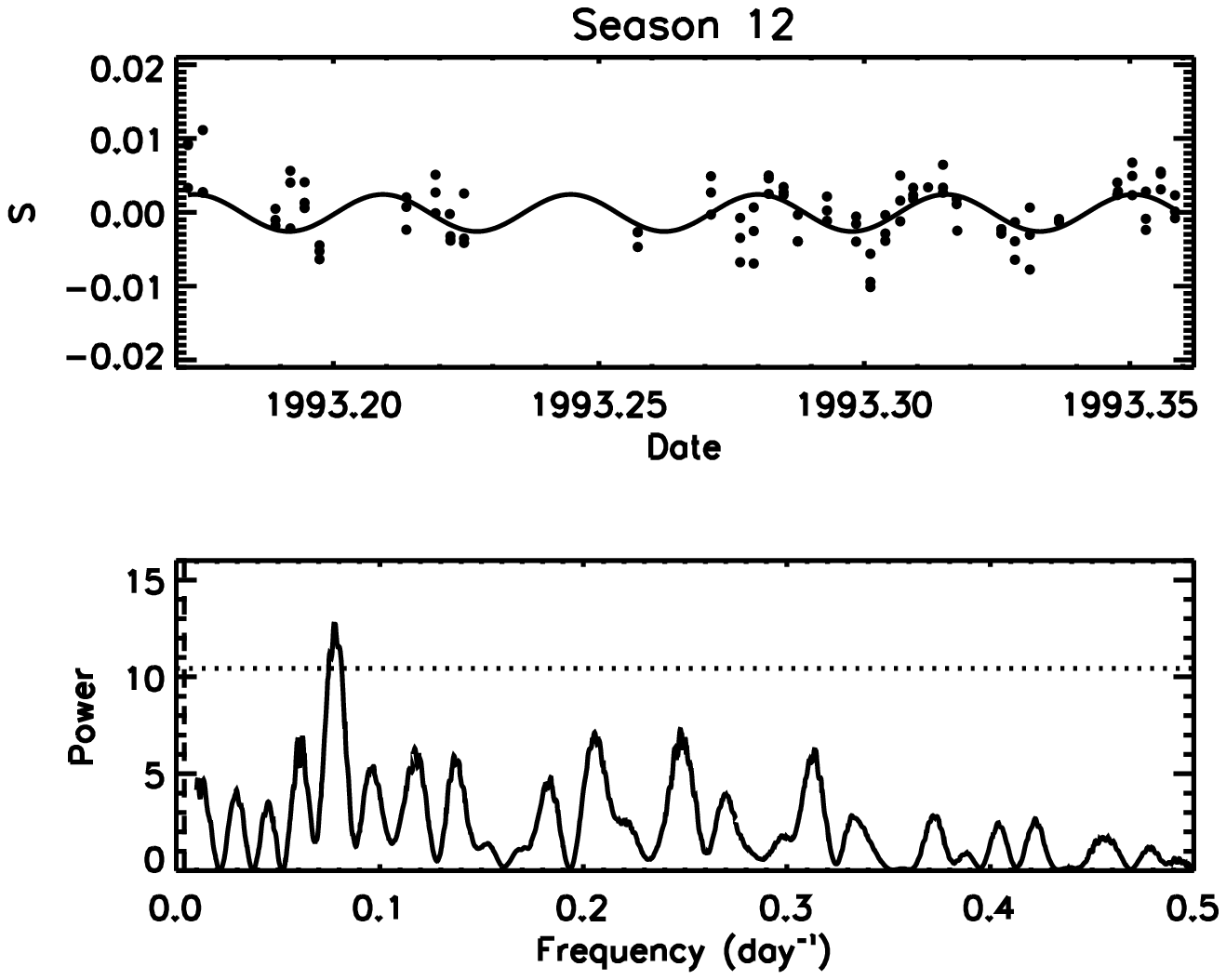}&
		\includegraphics[angle=0, width=80mm]{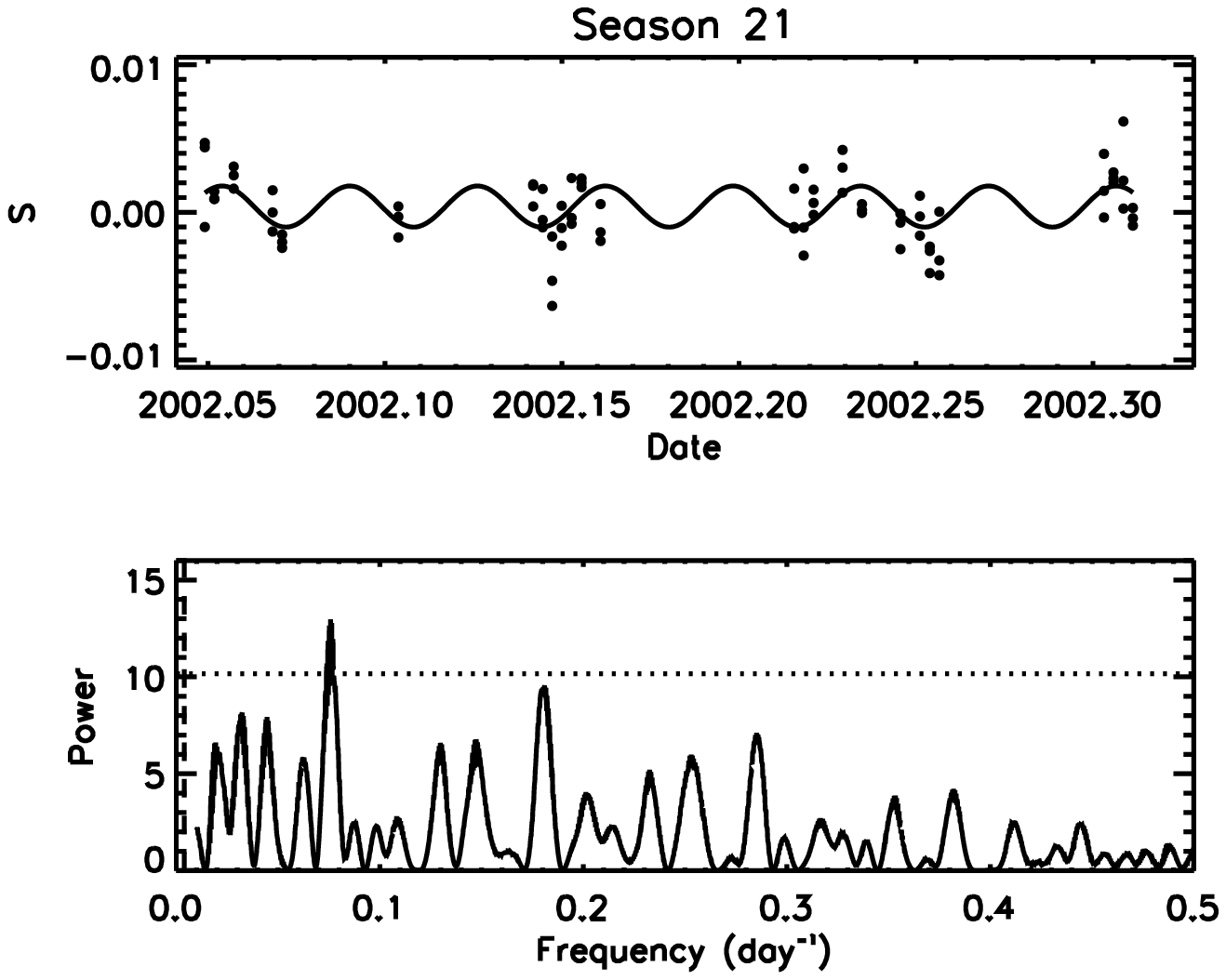}\\
	\end{tabular}
   \end{figure*}
   
       \begin{figure*}
    	\caption{One season shows a peak in HD 92788 that is consistent with rotation. It is graded as \textit{weak}.}
	\centering
    	\begin{tabular}{c}
     		 \includegraphics[angle=0, width=80mm]{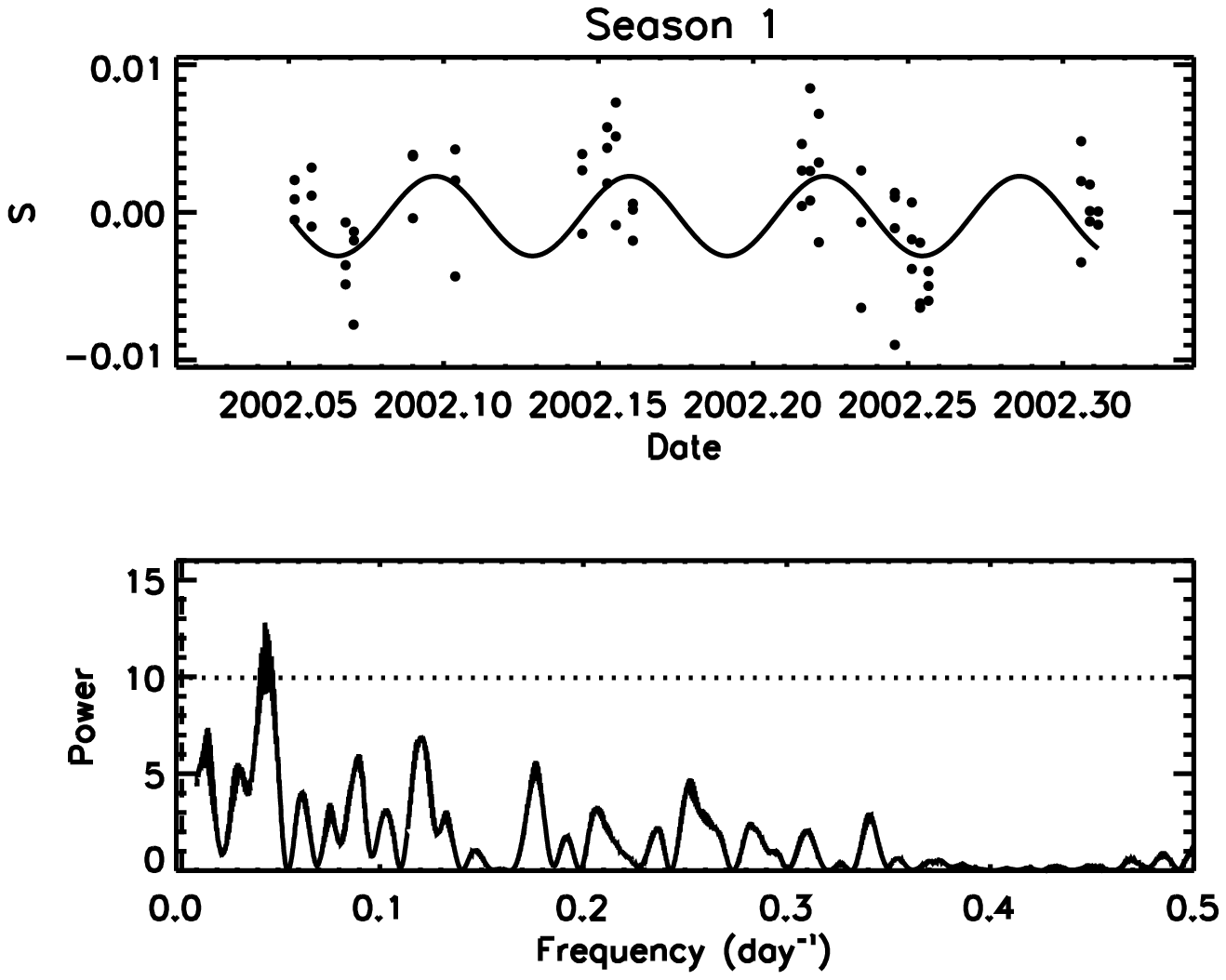} \\[5ex]
   	 \end{tabular}
   \end{figure*}
   
       \begin{figure*}
    	\caption{Photometric observations of HD 130322 reveal strong and persistent rotation modulation of the lightcurve by spots. The period is graded as \textit{confirmed}. }
	\centering
    	\begin{tabular}{cc}
     		\includegraphics[angle=0, width=80mm]{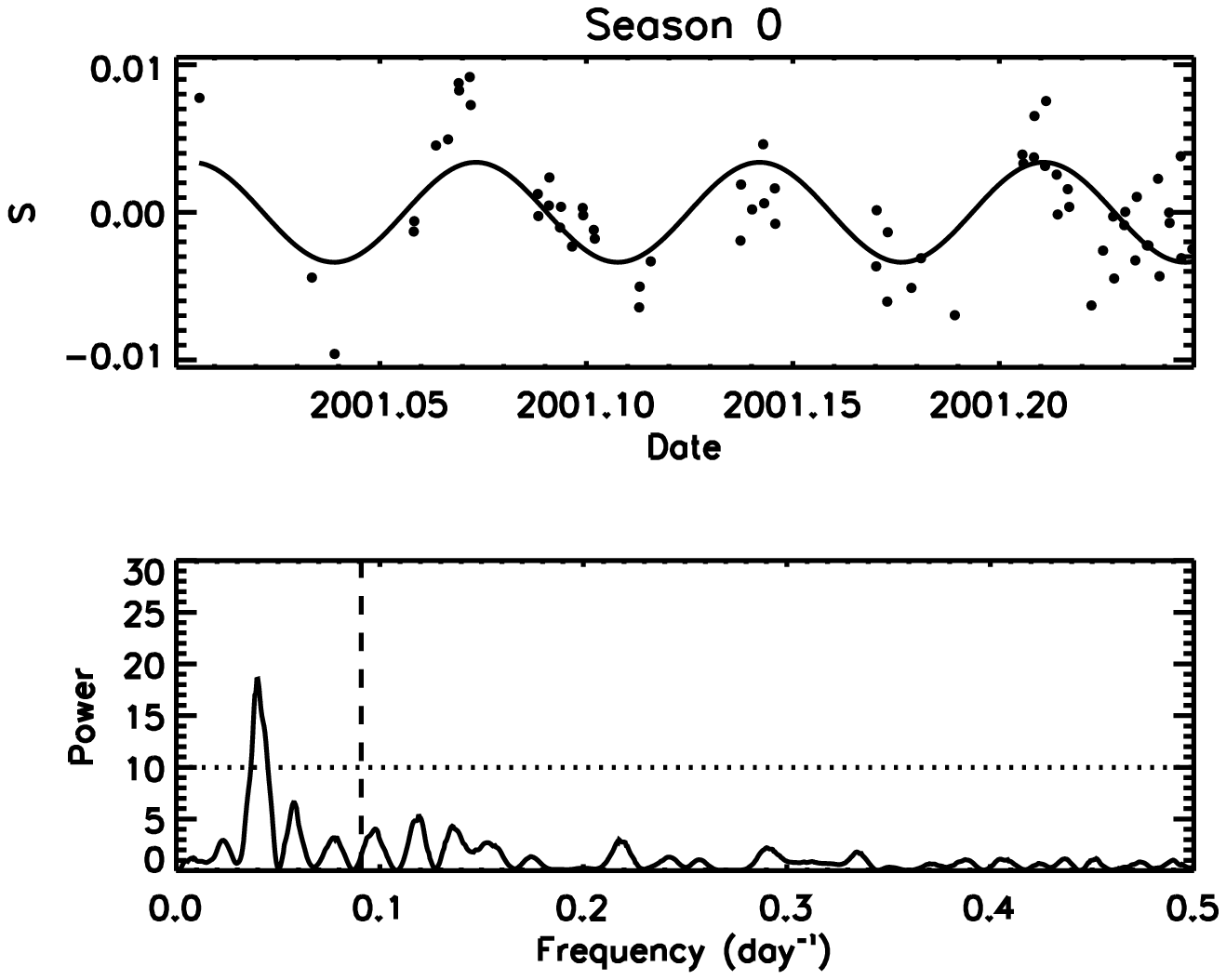} &
		\includegraphics[angle=0, width=80mm]{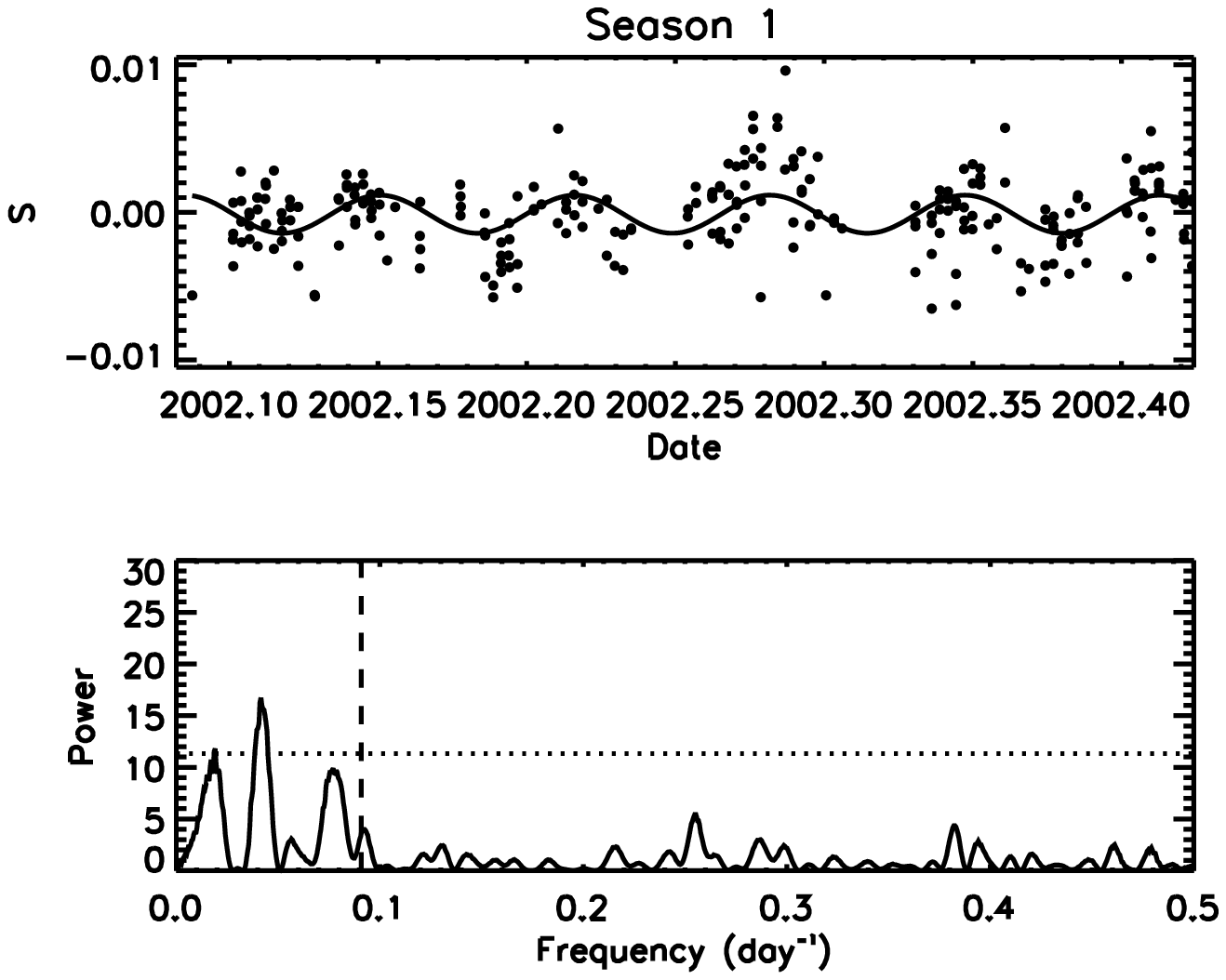}\\[5ex]
		\includegraphics[angle=0, width=80mm]{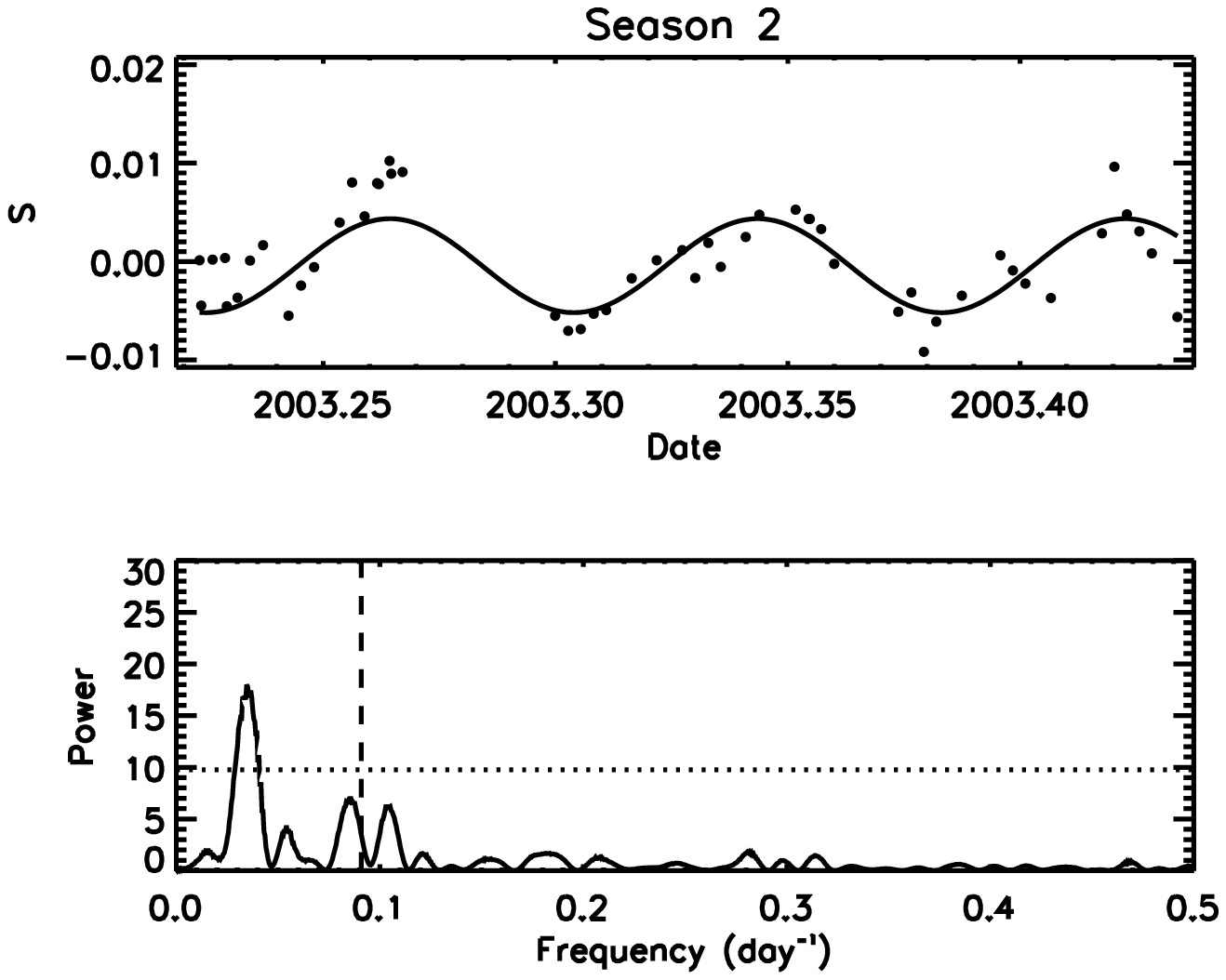} &
		\includegraphics[angle=0, width=80mm]{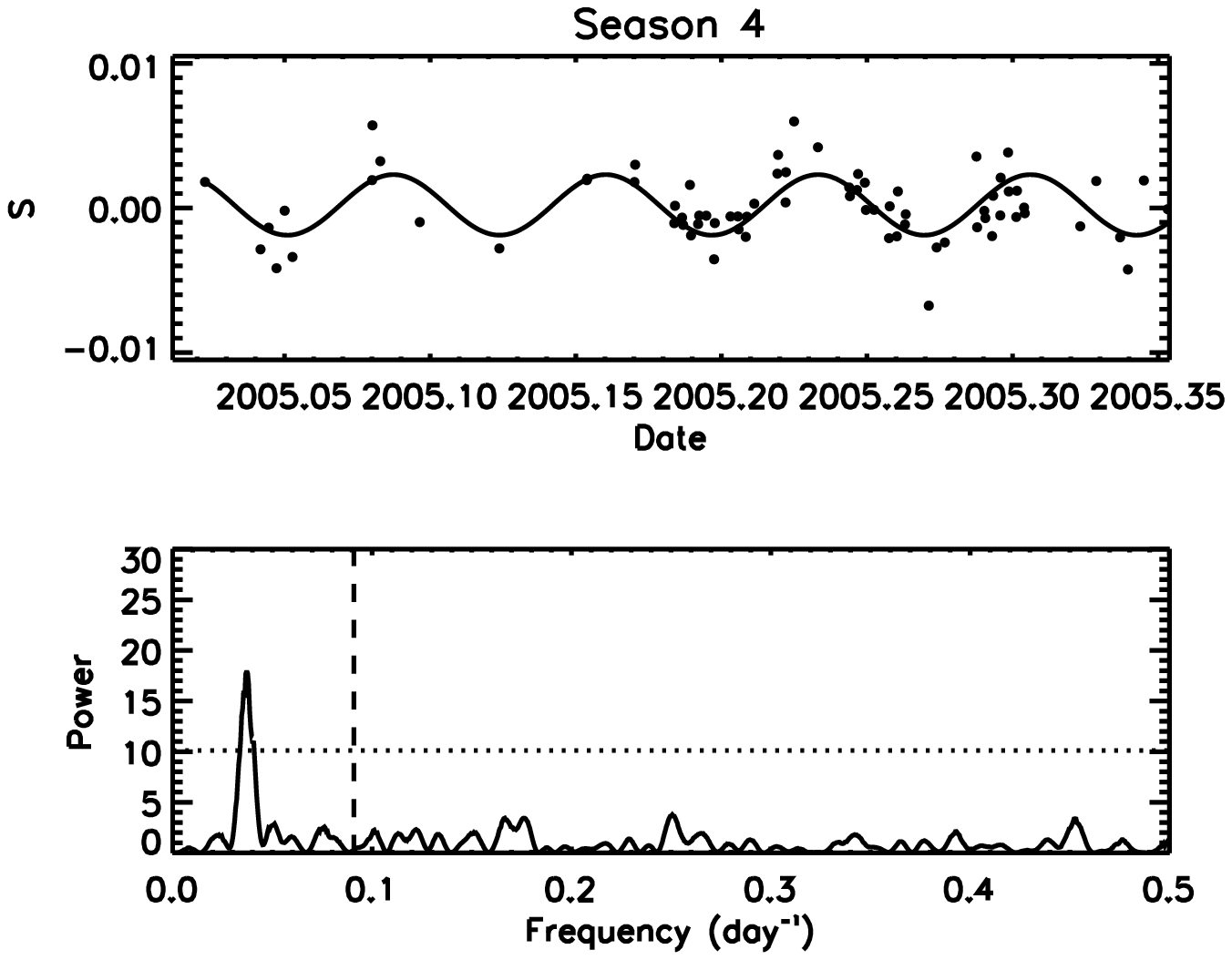}\\
	 	\end{tabular}
   \end{figure*}
   
          \begin{figure*}
    	\caption{Two seasons show peaks in the periodogram at the same period in HD 154345. The period is therefore designated as \textit{probable}. }
	\centering
    	\begin{tabular}{cc}
     		\includegraphics[angle=0, width=80mm]{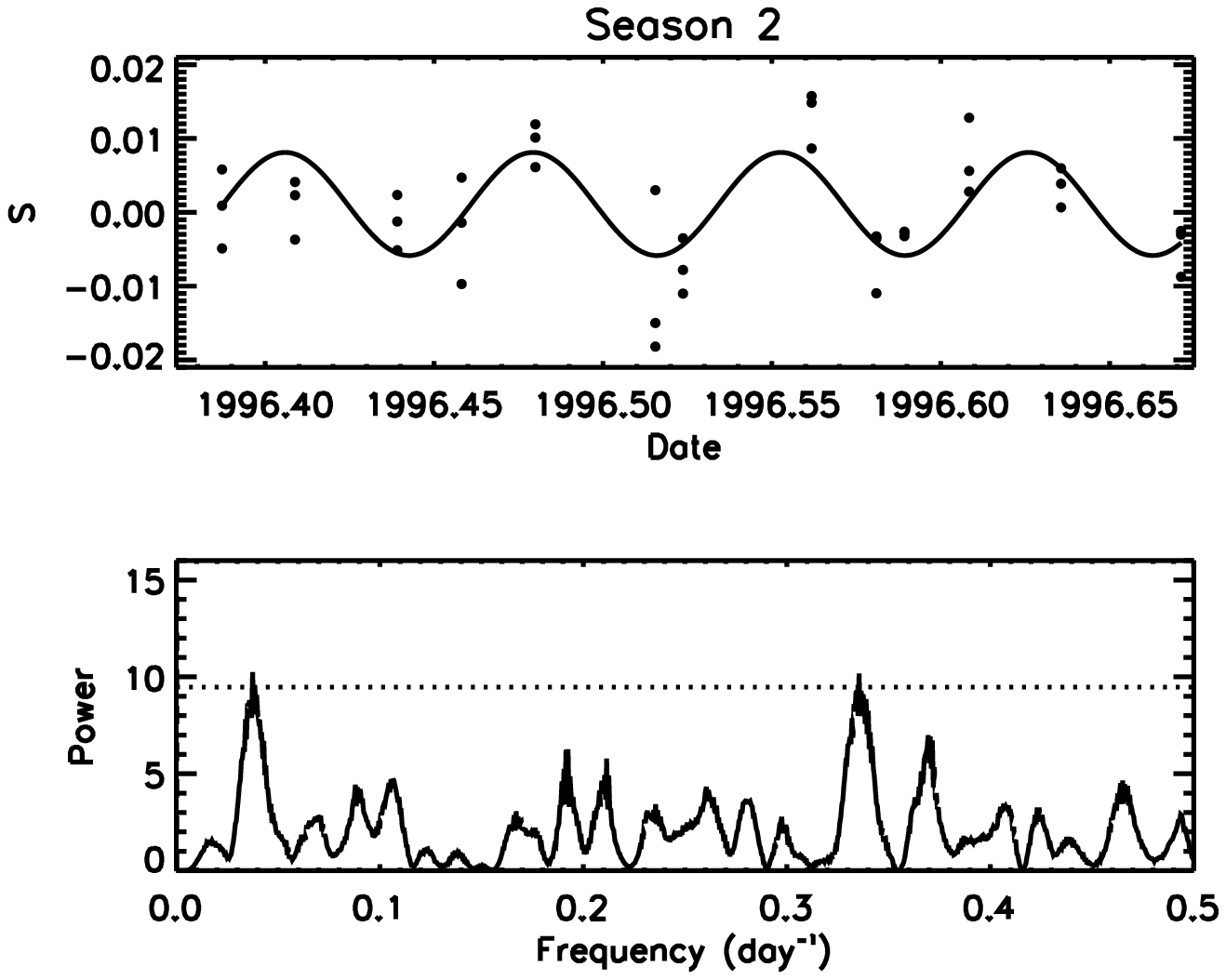} &
		\includegraphics[angle=0, width=80mm]{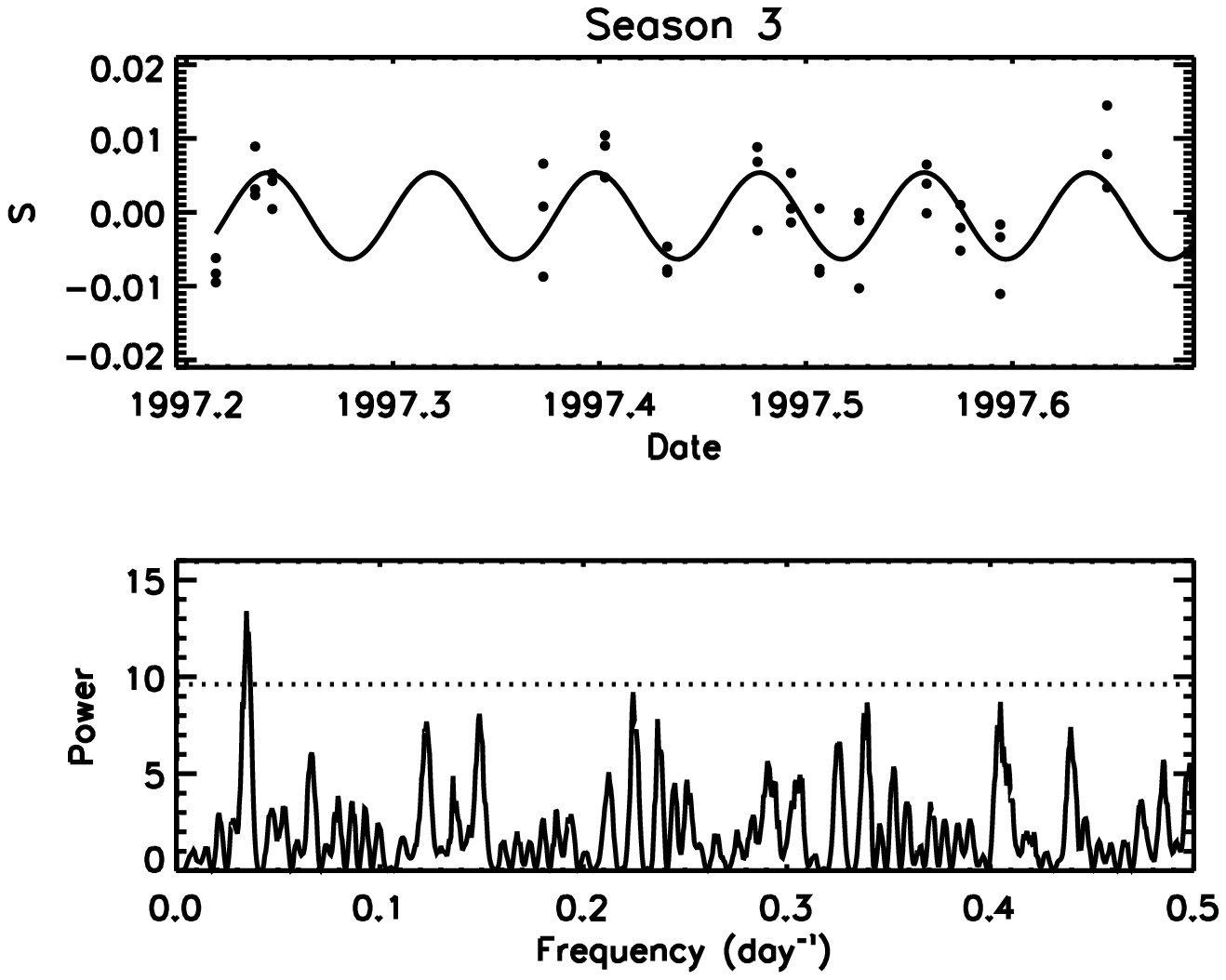} \\[5ex]
   	 \end{tabular}
   \end{figure*}
   
             \begin{figure*}
    	\caption{Only one period shows a period consistent with rotation in HD 217014 (51 Peg) however it is relatively strong so its designation is raised from \textit{weak} \citep{Henry00} to \textit{probable}. }
	\centering
    	\begin{tabular}{c}
		 \includegraphics[angle=0, width=80mm]{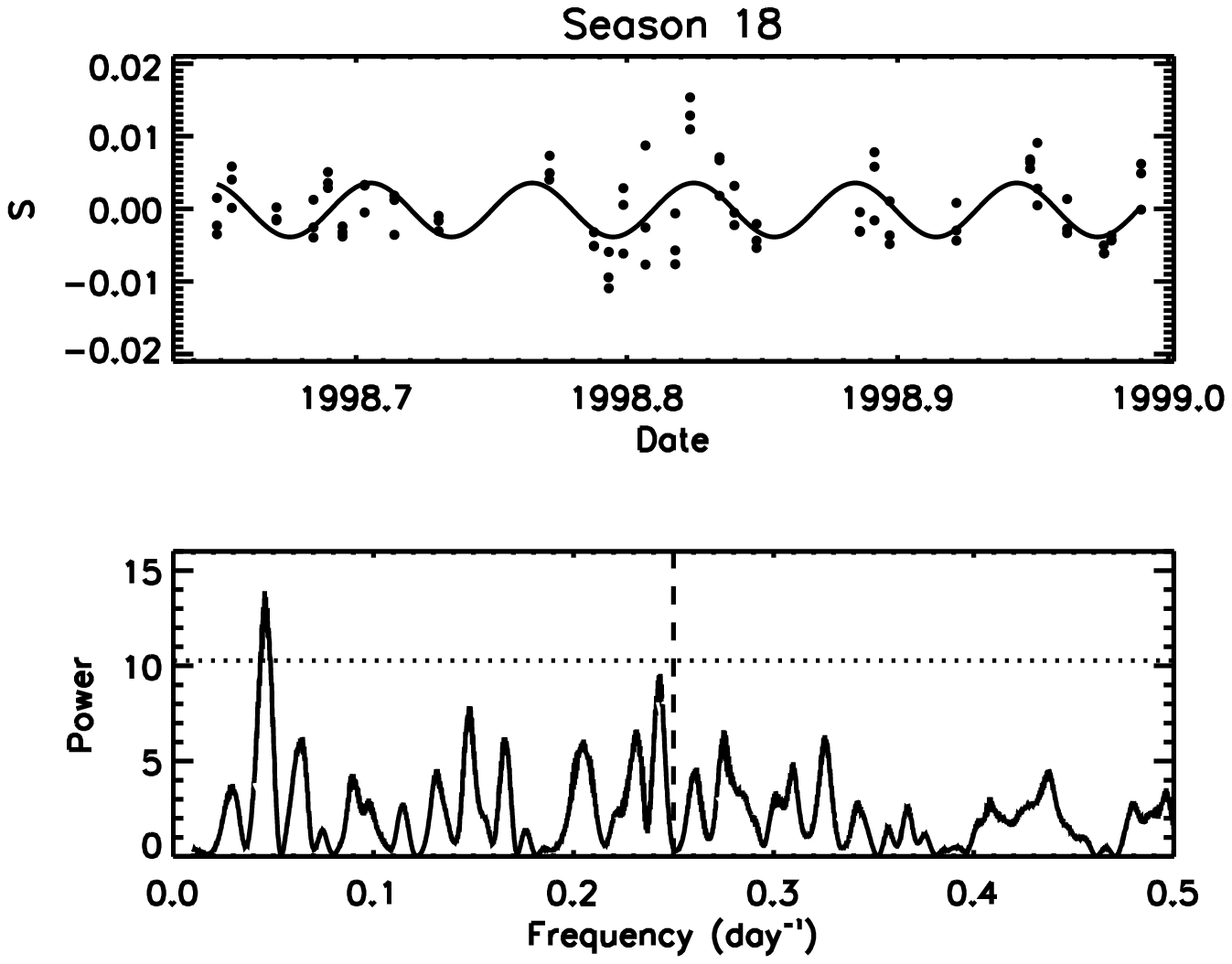} \\
   	 \end{tabular}
   \end{figure*}

\end{document}